\documentclass[
superscriptaddress,
 amsmath,amssymb,
 aps,
]{revtex4-2}

\usepackage{color}

\usepackage{ulem}
\usepackage{textgreek}
\usepackage{commath}
\usepackage{gensymb}
\usepackage{float}
\usepackage{graphicx}% Include figure files
\usepackage{dcolumn}% Align table columns on decimal point
\usepackage{bm}% bold math

\begin{document}

\preprint{APS/123-QED}

%\title{Spore Escape on Coherent Air Structures of a Fluttering Leaf}% Force line breaks with \\
\title{Droplet Outbursts from Onion Cutting}

\author{Zixuan Wu}
 \affiliation{%
Sibley School of Mechanical and Aerospace Engineering, Cornell University, Ithaca, NY 14853, USA.
}%%Lines break automatically or can be forced with \\


\author{Alireza Hooshanginejad}
 \affiliation{%
 Department of Biological and Environmental Engineering, Cornell University, Ithaca, NY 14853, USA.
}%%Lines break automatically or can be forced with \\

\affiliation{
Home Environment R\&D Department, SharkNinja Inc., Needham, MA 02494.
}

\author{Weilun Wang}
 \affiliation{%
Sibley School of Mechanical and Aerospace Engineering, Cornell University, Ithaca, NY 14853, USA.
}%%Lines break automatica

\author{Chung-Yuen Hui}
 \affiliation{%
Sibley School of Mechanical and Aerospace Engineering, Cornell University, Ithaca, NY 14853, USA.
}%%Lines break automatically or can be forced with \\

\author{Sunghwan Jung}%
 \email{sunnyjsh@cornell.edu}
\affiliation{%
Department of Biological and Environmental Engineering, Cornell University, Ithaca, NY 14853, USA.
}%

\date{\today}% It is always \today, today,

%airborne 
%\cite{ristaino2021persistent}

\begin{abstract}
\section*{Abstract}
Cutting onions often leads to tear-inducing aerosol release in kitchen, yet the underlying mechanics of droplet generation remain poorly understood. In this work, we combine custom-developed high-speed particle tracking velocimetry (PTV) and digital image correlation (DIC) to visualize and quantify droplet ejection during onion cutting. We show that droplet formation occurs via a two-stage process: an initial high-speed ejection driven by internal pressurization of the onion first-layer, followed by slower ligament fragmentation in air. By systematically varying blade sharpness and cutting speed, we find that faster or blunter blades significantly increase both the number and energy of ejected droplets. Strain mapping via DIC reveals that the onion’s tough epidermis acts as a barrier to fracture, enabling the underlying mesophyll to undergo significant compression before rupture, thereby increasing both the quantity and velocity of the resulting splashed droplets. Developing a scaling model and a simplified bi-layer model with a spring foundation, we experimentally and theoretically demonstrated how sharpened blades lead to not only fewer but also slower droplets. Numerical calculations accurately explain the onion critical fracture force obtained from independent Instron tests. The work highlights the importance of blade sharpening routines to limiting ejected droplets infected with pathogens in the kitchen, which pack additional outburst energy due to vegetables' outer strong casings.
\end{abstract}

%not used: and the stream flux Reynolds number as $\mathrm{Re}_{St}=0.12\mathrm{Re}_{b}$. We also discovered that

\maketitle

%\tableofcontents

\section{\label{sec:level1}Introduction}

Onion has been an essential agricultural product on the kitchen and medicinal tables since the ancient Egyptian and Sumerian ages in the 3000 B.C., or even earlier \cite{pareek2017onion,lanzotti2006analysis}. Ancient Egyptians worshiped onions believing that the concentric circles within symbolized eternity, and often placed them in the burial tomb of pharaohs \cite{pareek2017onion,lanzotti2006analysis, nmsuHistoryMexico}. However, throughout the span of culinary history, users have long been troubled by the lacrymatory effects induced by onion cutting and slicing, as Shakespeare described the discomfort in \textit{Antony and Cleopatra}: 
\textit{''Indeed the tears live in an onion that
should water this sorrow"}. When an onion is cut, the enzymatic reaction turns the sulfur compounds to propanethial S-oxide upon onion tissues rupture \cite{block2010garlic, kato2016production,spare1963lachrymatory}. When propanethial S-oxide reaches human eyes chemical activates the ciliary nerves in the cornea and cause tears \cite{spare1963lachrymatory}. Kohman first suggested that in addition to such chemicals being volatile, tiny hovering droplets containing them during cutting may be the true culprit, though evidence was lacking at his time \cite{kohman1947chemical}. However, there haven’t been any mechanistic review of droplet ejection in cutting onions supported by experimentation and visualizations.
%; Egyptians, and later Greeks and Romans consider onions to have tremendous health benefits, including aided endurance, reducing asthma, heart diseases, tumors, wounds, etcRecent studies indeed confirm its capability for inhibition of bone density loss
%and anti-coagulation \cite{mccallum2007onion}. 

%The scientific community has long assigned the lachrymatory factor in onion to propanethial S-oxide, which is produced by enzymatic reactions after tissue ruptures \cite{block2010garlic, kato2016production,spare1963lachrymatory}. Common perceptions of lachrymatory induction are chemical vapors containing the above chemical agents, catalyzed during cutting or crushing \cite{spare1963lachrymatory}. Kohman, in 1947, suggested in his \textit{Science} paper that tiny hovering droplets may be squirted from onion, instead of volatile fumes alone, though evidence was lacking at his time \cite{kohman1947chemical}. Past literature works thus have shown a lack of detailed mechanistic review of droplets ejection from onion cutting and visual inspection of the cutting sequence in higher spatial and temporal resolutions. Reducing high-speed ejections of minute onion droplets with respective to the cutting conditions can be a critical aspect for attenuating the tear induction effects. 

Coupling of fluids to elasticity in complex soft composites have shown rich emergent dynamics \cite{style2021solid,mehrabian2016soft, arnbjerg2024competition, plummer2024obstructed, lilin2024fracture, watson2018jet,wang2022effects,herrera2021spatiotemporal}. This applies also for fruits and vegetables. Smith et al. elucidated that microjets formations from the exocarps of citrus fruits depend on mechanical properties of the soft shell \cite{smith2018high}. Box et al. has shown that even a simple protective layer has rendered nuance stiffness changes for fruits \cite{box2020cloaking}. As biomimetic, soft material systems gain popularity for its often bio-compatibility, adaptability, and complex functionality \cite{shin2003biomimetic, trivedi2008soft, wallin20183d}, understanding how soft composites fracture and liquid inclusion exits pressurized complex soft composite can facilitate engineering designs. 

Beyond the scope of lachrymatory reduction and engineering concerns, onion cutting represents a common practice that lends well to the study of how kitchen cuts generate pathogenic splashes and contribute to the spreading of food-borne diseases. Atomized droplets are perfect carriers of viruses and bacteria, and there are many recent engineering designs aiming at reducing pathogenic splashes as the COVID pandemics has reasserted this notion \cite{wang2020covid,morawska2005droplet, garvey2023sink,hamidah2023spray,thurairajah2025splash}. In US alone, there are 76 million cases of food-borne diseases annually, most notably food poisoning by bacterial agents like \textit{Salmonella} and \textit{Campylobacter} \cite{newell2010food}. Malpractices in food preparation such as inadequate cutting board cleaning and the splashing of pathogen-carrying droplets are thought to play a major role as well \cite{klontz1995prevalence,lai2022handling,lai2023evaluation}. Does the cutting conditions in processing raw foods, such as speed and blade sharpness, matter in spawning such infected droplets? 

In this work, we tackled the above inquiries by using an experimental approach to investigate the combined mechanics of fluid fragmentation and soft tissue fracture at high spatial and temporal resolution, for the practice of onion cutting.

\begin{figure*}[h!tbp]
    \begin{center}
    \centering
  \vspace{-10pt}
    \includegraphics[width=1\textwidth]{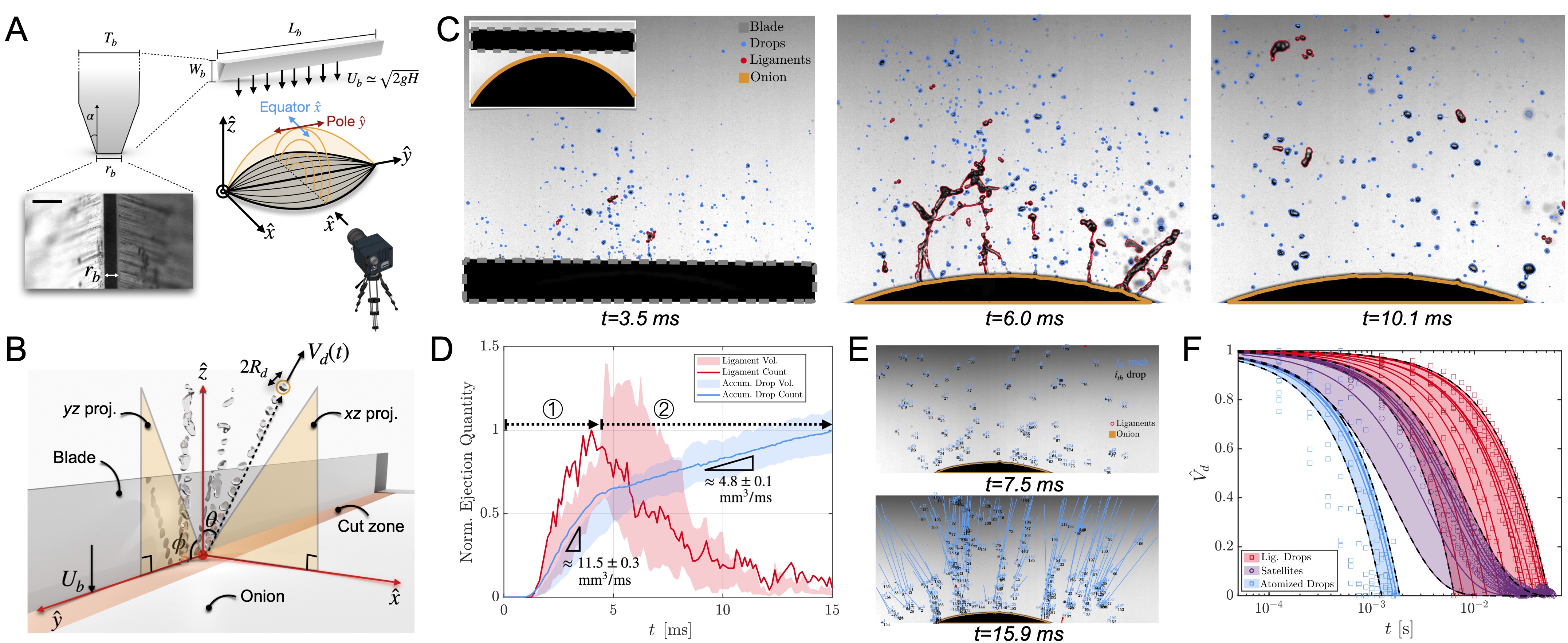}
    \caption{\textbf{Droplet release from onion cuts.} \textbf{A.} Guillotine experiment setup for capturing onion cutting with the onion, the blade, and the camera. Small inset is the SEM image of the blade tip. Scale bar is 20 $\mu$m. \textbf{B.} Visualization of the cutting, ejection, and rupture, with defined variables and geometry for the problem. \textbf{C.} Time sequence of the droplet ejection from an onion cut, at $t=$3.5 -- 10.1 ms; image segmentation divides detections into drops, ligaments, and macro objects of onion and blade. $r_\mathrm{b}\approx 13\ \mu$m and $U_\mathrm{b}\approx 2.1$ m/s. An inset in the first panel shows the outline of onion and the blade before the cut. See movie in SI Movie S1. \textbf{D.} Ejection metrics with max value aligned at 1 over time. Metric 1: mean value of counts (lines); metric 2: mean value of volumes with standard error bands (shades). Ejection type: ligament (red); accumulated drops (blue), meaning drops ejected accumulated from $t=$0.0 ms. Two different regimes with different ejection rates and mechanisms are revealed. Averaged from 7 samples with different cutting speeds. \textbf{E.} Demonstration of customized particle tracking velocimetry for associating drops and tracks in a frame, for $t=$7.5 -- 15.9 ms, with the same cutting parameters as \textbf{C}. See PTV movie in SI Movie S2. \textbf{F.} Normalized (from max to min velocity) decay profiles of drops from ligaments, satellite drops, and fast atomized drops, with the associated decay prediction in lines. Shade bands denote the range found for decay curves with the measured initial velocity and sizes.}
    \label{o1}
    \end{center}
\end{figure*}

\section{\label{sec:level1}Results}
\subsection{\label{sec:level2} Experiments}
We built a guillotine-style setup consisting of a vertical slider with a blade attached at the bottom to control the impact conditions between an onion hemisphere and a thin steel blade. The blade was released from a height $H$ to achieve free-fall cutting. The blade had a high aspect ratio, $L_\mathrm{b}/W_\mathrm{b} \gg 1$, where $L_\mathrm{b}$ and $W_\mathrm{b}$ are the length and height of the blade, respectively. The height $W_\mathrm{b}$ was 5 mm to enable visualizations near the onion surface. The blade sharpness was characterized by the tip radius $r_\mathrm{b}$, measured using scanning electron microscopy (SEM), while other geometric parameters, including the wedge angle ($\alpha \approx$8.5$\degree{}$), were kept constant. 

Fig.~\ref{o1}\textit{A} shows the experimental setup with the $\hat{y}$ axis defined along the onion's pole direction, i.e. from root to shoot. A representative SEM image of the blade tip radius $r_\mathrm{b}$ is shown in the inset. More details on sample preparation are provided in Supplementary Note 1. Cutting speeds is on the order of 0.1 to 1.0 m/s, measured from video records of onion cutting by chefs on public records (see Supplementary Note 1). A schematic is shown in Fig.~\ref{o1}\textit{B}, where key variables are defined. The primary observational plane corresponds to the $yz$ projections of the rupture. Droplet diameter/radius and velocity are denoted as $D_\mathrm{d}$, $R_\mathrm{d}$ and $V_\mathrm{d}$, respectively. To minimize the $yz$ projection error in the droplet speed, the depth of field was limited to 1 mm. For analysis, we approximate $V_\mathrm{d}$ from $V_\mathrm{{d,z}}$, the $yz$ projection of the velocity. Error estimation procedures are described in the Methods section. High-speed videos at 5,000 -- 20,000 frames per second (fps) show that droplets are ejected from the fracture site immediately upon onion rupture, with an initial velocity of 1.0 -- 40 m/s. This ejection rapidly fragments into hundreds to thousands of satellite droplets within 0 -- 20 ms as shown in Fig.~\ref{o1}\textit{C}. The full sequence is available in Supplementary Fig. 1 and SI Movie S1. 

%Nonetheless, the nonzero width of the blade still presents a challenge, which is the blockage of view for liquid release right at the opening, and a potential complication, which is the hydrophilic adhesion of the ligaments on the blade surface. Resolution for the blockage issue is still currently missing, and we compromised by back-calculating the initial velocity from early-stage measurements. As for the complication, regular stainless steel kitchen knifes are hydrophilic, which means most of the ligaments ejected directly upwards would attach to the blade and produce no droplets. However, as the following sections would suggest, there is a distribution on the angle that the ligaments eject vertically at, away from 90$\degree$. Therefore, we attempt to limit the analysis here primarily to ligaments and breakup occurring without blade attachments. 

As shown in Fig.~\ref{o1}\textit{C}, the ejected liquid emerges as either distinct droplets or ligaments, followed by successive breakup processes. By analyzing the ejected liquid across all frames, we extract temporal data on the count and volume of both ligaments and droplets. Fig.~\ref{o1}\textit{D} shows the ejection of accumulated drops and ligaments over time, revealing two distinct stages of droplet formation. The first stage is the initial violent rupture, akin to industrial atomization, in which the fluid stream is rapidly disintegrated into fine droplets \cite{lefebvre2017atomization}. This stage exhibits a high drop generation rate of 11.5$\pm$0.3 mm$^3/$ms. The second stage results from ligament fragmentation, which begins with a gradual decay in ligament counts and volume as they break apart and get converted into droplets, leading to a lower but steady drop generation rate of 4.8$\pm$0.1 mm$^3/$ms. 

Hydrodynamics instabilities during primary breakup from atomization to wind-induced fragmentation are observed, which is driven by interactions among capillary, inertial, and aerodynamics forces \cite{lin1998drop, dumouchel2008experimental}. The onion juice is measured to be a shear thickening fluid in the typical cutting speed regime, with viscosity $\mu_\mathrm{d}$ at 0.002-0.005 Pa$\cdot$s, surface tension $\gamma_\mathrm{d}\simeq$52.5 mN/m. Due to the high cellular water content, density $\rho_\mathrm{d}$ is approximated at $\rho_\mathrm{d}\simeq \rho_\mathrm{d, water}=$1000 kg/m$^3$. A detailed discussion of breakup processes and onion juice properties is provided in Supplementary Note 1 and Supplementary Fig. 2. 

%Another type of droplets observed are satellite droplets, which are produced via secondary disintegration, from both initial ruptures and ligament breakup. \red{Not clear the difference between the primary and secondary breakup processes.} These droplets are typically much smaller in size, have lower ejection velocities, and tend to drift under ambient conditions. 

To analyze droplet trajectories and statistics, we used a custom particle tracking velocimetry tool (PTV), suitable for detecting small droplets and developing cross-frame linkages. Implementation details are provided in Supplementary Note 2 and Supplementary Algorithms 1 and 2. Tracking results are shown in Fig.~\ref{o1}\textit{E}, where the $i^{th}$ droplet is linked to the $j^{th}$ trajectory, as labeled for a cutting sequence at $t=$7.5 -- 15.9 ms. Straight linkages across multiple frames indicate reliable track reconstruction, see Supplementary Fig. 3 and SI Movie S2 for the full sequence. However, the current tracking scheme is still prone to errors at higher droplet densities and speeds during early-stage ejections. 

Following ejection, droplet velocities in air rapidly decrease due to the combined effects of aerodynamic drag and gravity. 
The temporary decay of velocity in the $\hat{z}$ direction can be described as \cite{kim2019vortex,smith2018high} 
\begin{equation}  \label{dvdt}
\frac{dV_{d}}{dt}=-\frac{3 \alpha \pi \mu_a D_\mathrm{d}}{m_\mathrm{d}}V_{d}-g.
\end{equation}
where $m_\mathrm{d}$ is the drop mass, and $\mu_a$ is the dynamic viscosity of air (0.017 cP). The coefficient $\alpha$ is a drag correction factor as $\alpha=1+0.15\mathrm{Re}_\mathrm{a}^{0.69}+0.018\mathrm{Re}_\mathrm{a}(1+4.3\times 10^4\mathrm{Re}_\mathrm{a}^{-1.2})^{-1}$. This expression is valid for subcritical Reynolds numbers $\mathrm{Re}_\mathrm{a}= \rho_\mathrm{a} U_\mathrm{d} D_\mathrm{d}/\mu_\mathrm{a}$ below 10$^5$, \cite{smith2018high,clift1971motion}. 

Fig.~\ref{o1}F presents the normalized velocity decay profile over time, defined as $\hat{V}_\mathrm{d}=(V_\mathrm{d}-V_\mathrm{0})/(V_\mathrm{f}-V_\mathrm{0})$, where $V_\mathrm{0}$ and $V_\mathrm{f}$ are the initial and final in-frame velocities, respectively. This normalization enables comparison across droplets from a wide range of initial velocities with the numerical calculations. For the three types of droplets aforementioned, we extracted individual trajectories and overlaid corresponding decay predictions. Shaded regions indicate the prediction envelopes based on given $V_0$ and $R_\mathrm{d}$. The observed decay profiles mostly match predictions for ligament breakup, while smaller satellite and atomized droplets show greater deviations from predictions. This is presumably due to uncontrolled ambient airflow caused by the falling blade.

With reliable PTV tracking and a clear understanding of ejected drop dynamics, we turn to examine the influence of two cutting parameters: cutting speed $U_\mathrm{b}$ and blade sharpness $r_\mathrm{b}$.

\begin{figure*}[h!tbp]
    \begin{center}
    \centering
  \vspace{-10pt}
    \includegraphics[width=1\textwidth]{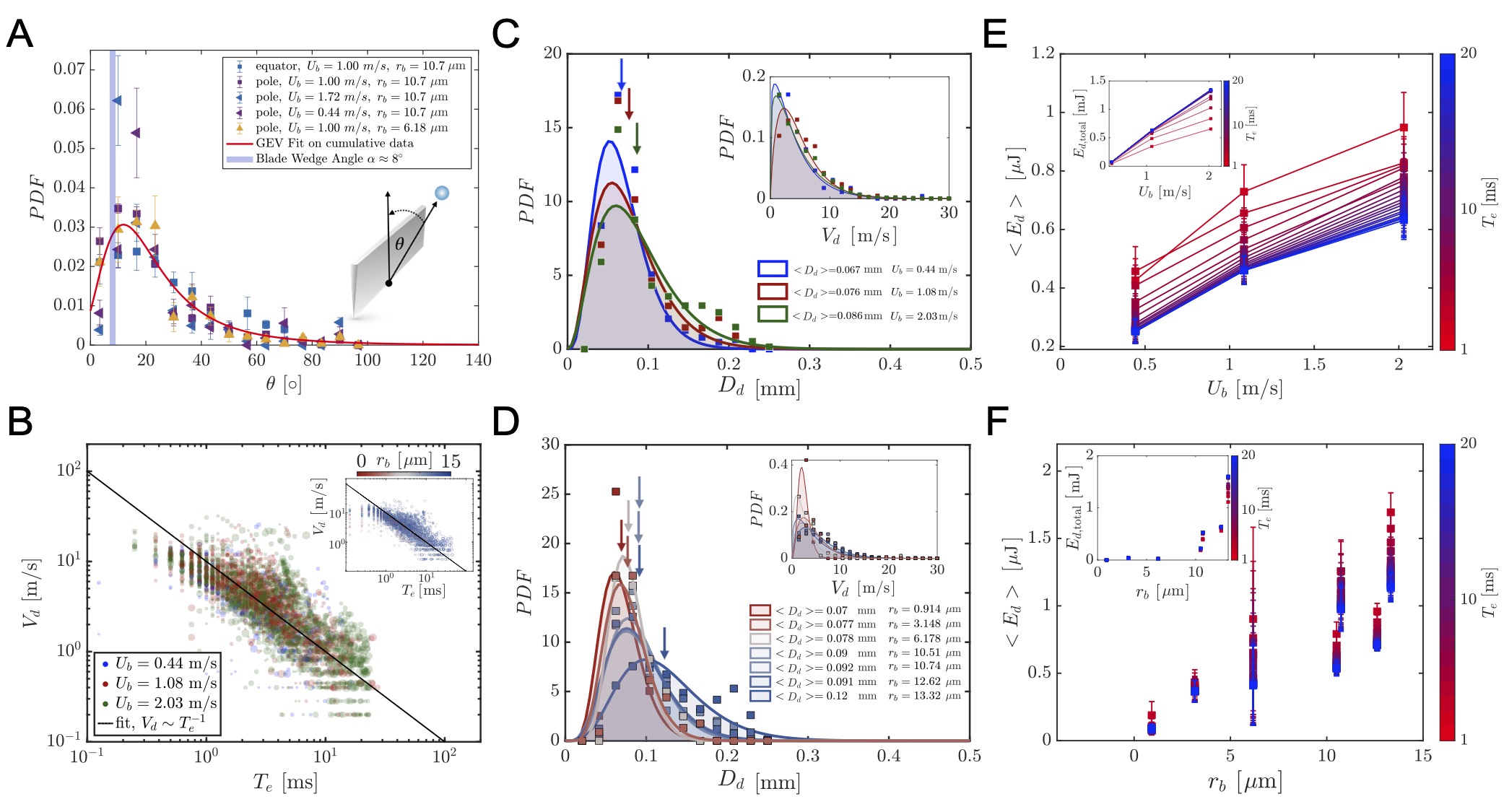}
   \caption{\textbf{Droplet distribution and energetic characterizations.} \textbf{A.} For different combinations of cutting orientations (\textit{equator} or \textit{pole}), cutting speed ($U_\mathrm{b}$), and blade sharpness ($r_\mathrm{b}$), the angles are extracted from side-view visualizations of the droplet tracks. A fit of generalized extreme values is applied to the cumulative data. The wedge angle of the blade used is measured at 8$^{\degree{}}$, with dental polymer impression technique. \textbf{B.} Droplet velocity, $V_\mathrm{d}$ as a function of ejection $T_\mathrm{e}$ for different velocities and different blade sharpness. Fits indicate a relation of $V_\mathrm{d} \propto T_\mathrm{e}^{-1}$. \textbf{C-D.} Probability density functions of drop diameters $D_\mathrm{d}$ and velocities $V_\mathrm{d}$ (inset) of droplets ejected, for different impact velocity $U_\mathrm{b}$ at $r_\mathrm{b} \simeq$10.5 $\mu$m (\textbf{C}) and different blade sharpness $r_\mathrm{b}$ at $U_\mathrm{b}\simeq$1.17 m/s (\textbf{D}). Arrows label the mean value of the respective distribution. \textbf{E-F.} Average energies and total energies (inset) of droplets color coded for ejection time, for different $U_\mathrm{b}$ \textbf{(E)} and sharpness $r_\mathrm{b}$ \textbf{(F)} from the sets of data in \textbf{C-D}. }
    \label{o2_2}
    \end{center}
\end{figure*}

\subsection{\label{sec:level2} Statistics and Energetics}

The overall ejection pattern is predominantly confined to a single plane, oriented nearly parallel to the blade's penetration plane in the $\hat{z}$ direction as expected since the compression is vertical. The distribution of ejection angles $\theta$ as defined in Fig.~\ref{o1}B, is shown in Fig.~\ref{o2_2}A and consistently peaks around 10$\degree$. This slight deviation from a purely vertical release is likely attributed to the wedge angle of the blade, which is 8.5$\degree$. Mann-Whitney U-tests are conducted to assess shifts in highly skewed distributions. No statistically significant differences in median ejection angle at the 5\% significance level are found across different $r_\mathrm{b}$, $U_\mathrm{b}$, or cutting orientations. Blade sharpness is the only parameter that causes a statistically significant shift in the ejection angle with $p=$ 0.017, possibly due to the sharpening process slightly altering the wedge condition. Details on the sharpening procedure are explained in the Methods section. 
%Both orientation and velocity changes yield $p=$0.095; only the sharpness group is the only one with successful rejection with a $p=$0.017.

Pressure within the onion layer during blade penetration dictates the speed of rupture. Assuming that the internal pressure does not get relaxed significantly at the moment of fracture, the liquid flow from the pressurized region to the ejected ligaments can be approximated by the scaling relation of $P_\mathrm{e}\sim \rho_\mathrm{d} V_\mathrm{d,0}^2$ \cite{smith2018high,hasegawa1997anomaly}. Here, $V_\mathrm{d,0}$ is the initial droplet speed. Fig.~\ref{o2_2}B shows that $V_\mathrm{d,0}$ decays with ejection time $T_\mathrm{e}$ with a power law with exponent $n$=--1, regardless of the parameters. This decay profile over two orders of magnitude implies that fluid originating from deeper within the tissue experiences higher dissipation before reaching the surface. As a result, the fastest droplets are ejected during the initial rupture phase, posing the greatest risk of reaching the cutter's eyes. We further observe that the total volume of ejected droplets with both blade sharpness $r_\mathrm{b}$ and cutting speed $U_\mathrm{b}$, indicating that these parameters influence the volume of tissue affected during cutting. 

%Correlating ejection speed with pressure $P_{cell}$ at release, the static and dynamic pressure balance can be calculated as $P_{cell}=0.5\rho_\mathrm{d}V_\mathrm{d,0}^2+\Gamma$, where the second term accounts for viscous loss from orifice constriction, approximated by $\Gamma\approx 1.25\rho_\mathrm{d}V_\mathrm{d,0}^2$ \cite{smith2018high,hasegawa1997anomaly}.The plateau value of max velocity is reached at smaller ejection time; this implies the most pressurized cuts and ejection occurs with the first layer, resulting in spray-like ejections as shown in Fig.~\ref{o2_1}D. Such plateau of max speed appears to be increasing with blade sharpness. Surface tension is ignored here since the Weber number is above 300 for most of the atomized droplets from first layer.

The probability density functions (PDFs) of drop diameter ($D_\mathrm{d}$) and initial velocity ($V_\mathrm{d,0}$) both follow a Gamma-type distribution of the form as
\begin{equation} \label{dist}
P(x=D_\mathrm{d}, V_\mathrm{d})=\frac{1}{b^a \Gamma(a)}x^{a-1} e^{-x/b},
\end{equation} 
as shown in Fig.~\ref{o2_2}C-D. The number of ejected droplets increases by around fourfold as the cutting speed $U_\mathrm{b}$ increases from 0.44 to 2.03 m/s, and up to forty-fold as blade sharpness $r_\mathrm{b}$ increases from 0.91 to 13.3 $\mu$m, as shown in Supplementary Fig. 4. For onion ejections, the fitting parameters change the ranges of $a=$2.0 -- 12 and $b=$0.0050 -- 2.0. Mann-Whitney U-tests for a no-difference null hypothesis shows statistically significant droplet sizes increments with increasing cutting speeds, with $p\leq$0.015 comparing low to high cutting speeds. %\red{Is it correct?}. \blue{sorry, what is? I checked. the drop height of 21 cm yields to 2 m/s and the p value is correct, it is very low, meaning rejection of null. } \red{I clarify it at the end of this paragraph.} 
However, the statistical results for $V_\mathrm{d}$ variations are less clear, which suggests that the pressurization and rupture in soft tissue are more dependent on material and blade properties. %\red{Here, you said $D_d$ increases with $U_b$, but $V_d$ is not correlated with $U_b$ very well. Then, your last sentence says that these suggest pressurized ruptures. This part was not clear to me.  }\blue{I reworded these. please check. }

Blade edge width $r_\mathrm{b}$, on the other hand, shows a statistically significant positive correlation with both droplet size and speed. Compared to the sharpest blade ($r_\mathrm{b}=0.91\ \mu$m), p-value is from 0.1 to 0.2 for similarly sharp blades, showing no significant difference. However, for blades with $r_\mathrm{b}>7\ \mu$m, p-value drops below 0.01, which indicates a statistically signifcant increase in droplet size. Similarly, $V_\mathrm{d}$ scales positively with increasing $r_\mathrm{b}$ with p-value consistently below 0.01 comparing with the sharpest blade.

The Gamma-type distribution for sizes is consistent with previously observed distributions in ligament breakup processes  \cite{villermaux2007fragmentation}. The distribution results from a mixture of ligaments and droplets in the frame, %\blue{please check if this makes sense now.}
and the thick tail of the skewed distribution is a consequence of droplet coalescence \cite{villermaux2007fragmentation}. 
% \red{Do people say "an exponential size fall off"? Or should we call "a thick long tail"?} \blue{I reworded this. please check.}
%\red{Are you referring to the thick tail in PDF? I don't see them very well in C and D.} \blue{yes, it just means the tail, this is a description of the gamma from literaure, we can exclude if you deem it unimportant }. 
Velocity distributions, on the other hand, do not follow a previously reported Gaussian distribution for unsteady sheet fragmentation \cite{wang2018unsteady}. The breakup dynamics here exhibit signature of first and second wind-induced instabilities with violent shearing of the liquid film and highly irregular ligament shapes, which differ remarkably from the scenario of an impinging drop expansion with regular rim pinch-off \cite{wang2018unsteady}.

To further understand the energy balance in the cutting process, we conduct the energy analysis by dividing the blade's input energy into the energy spent on cutting (fracture and dissipation) and the energy associated with drop ejections as
$E_\mathrm{b}=m_\mathrm{b} U_\mathrm{b}^2/2=E_\mathrm{\mathrm{cut}}+E_\mathrm{d,total}$, where $E_\mathrm{d,total}=\sum{m_\mathrm{d}V_\mathrm{d}^2/2}$ represents the total kinetic energy of the ejected droplets. The average droplet energy, $\langle E_\mathrm{d} \rangle$, is the total kinetic energy divided by the number of the ejected droplets. The averaged droplet energy positively correlates with the blade speed, as shown in  Fig.~\ref{o2_2}E. The inset also reveals a similar monotonic increase in $E_\mathrm{d,total}$. Time-based color coding further suggests that the average drop energy is highest immediately following ejection and decays over time. Assumming an equal partitioning between between cutting and drop ejection, one expects $E_\mathrm{d,total}\sim U_\mathrm{b}^2$. However, the creation of a new fracture surface and energy dissipation in the onion tissue both depend on cutting speed, due to the viscoelastic nature of vegetables and fruits \cite{yamagata2017experimental,mahiuddin2020application}. As a result, we observe a sub-linear scaling of droplet energy with increasing $U_\mathrm{b}$, reflecting the nonlinear increment of $E_\mathrm{cut}$ with cutting speed $U_b$. 

As $r_\mathrm{b}$ increases, the droplet energy increases monotonically, as shown in Fig.~\ref{o2_2}F. In particular, the total droplet energy $E_\mathrm{d,total}$ scales with $r_\mathrm{b}$  superlinearly, mostly due to the increase in ejected volume, expected from scaling relation of $E_\mathrm{d,total}\sim r_{\mathrm{plastic}}^3$, where $r_{\mathrm{plastic}}$ denotes the plastic fracture zone as the blade passes through the onion, which itself increases with size of the blade $r_\mathrm{b}$. 

However, the average $V_\mathrm{d}$ also increases with larger $r_\mathrm{b}$. The pressurization beneath the blade should inversely scale with $r_\mathrm{b}$ instead, due to the reduced stress concentration and contact pressure. Nonetheless, Fig.~\ref{o2_2}D and F show that droplet diameter $D_\mathrm{d}$, velocity $V_\mathrm{d}$, and average energy $\langle E_\mathrm{d} \rangle$ all increase with $r_\mathrm{b}$, especially during the early ejection stage ($T_\mathrm{e}<$ 1 ms) of the outermost layer of the onion. 
To elucidate blade-onion interactions, the following sections focus on how increasing $r_\mathrm{b}$ affects tissue compression and model subsequent pressurization during the early stage of blade-onion contact. In this early stage, droplets experience minimal dissipation and have the highest potential for spreading.

%and the distributions in terms of size and velocity with regards to cutting speed and sharpness. 

%then presents the different forces and the map. 
%depending on flow parameters is coordinated by the balance between capillary, inertial, and aerodynamics forces \cite{lin1998drop, birouk2009liquid}

\begin{figure*}[h!tbp]
    \begin{center}
    \centering
  \vspace{-10pt}
    \includegraphics[width=1\textwidth]{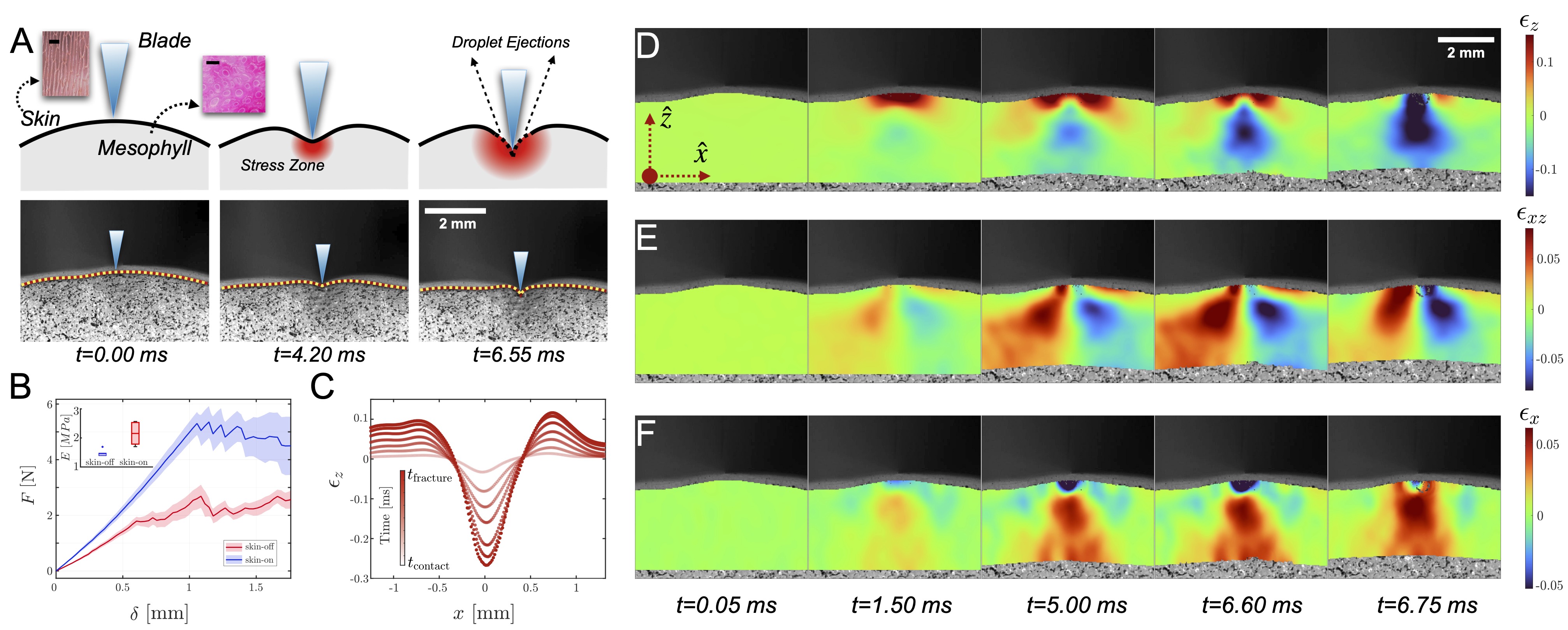}
    \caption{\textbf{High-speed digital image correlations on cutting cross section.} \textbf{A.} Schematics of the first-layer compression by a blade, composed of a epidermis and mesophyll layer. A stress zone is developed ahead of the blade tip, with significant deterrence by the skin layer. Droplets are ejected after the skin layer is broken through. The lower panels showcase the corresponding experimental images of the surface displacement during cutting of the first layer. See movie in SI Movie S3. Microscopy images showcase the different morphology of the epidermis (skin) versus mesophyll tissues (bulk). The scale bars are at 100 $\mu$m. \textbf{B.} The fracture force and displacement curves measured from a cylindrical flat punch geometry with an average of 5-6 samples for each skin-on/off condition. Inset shows the elasticity extracted from the linear portion. \textbf{C.} $\hat{z}$ direction strain $\epsilon_z$ for a horizontal line profile across the sample indicated in \textbf{A} from contact to fracture time. The data is one representative case, not the same sample in \textbf{A}. \textbf{D-F.} Time sequences of DIC extracted strain fields for the onion cross section shown in \textbf{A}: $\epsilon_z$ (\textbf{D}), $\epsilon_{xz}$ (\textbf{E}), and $\epsilon_x$ (\textbf{F}). See SI Movie S4.}
    \label{o3}
    \end{center}
\end{figure*}

\subsection{\label{sec:level2} Pressurization Mechanics}

%In addition, examining ligaments' initial jetting condition requires simultaneous time and space resolutions on the the order of of 10 $\mu$s and 10 $\mu$m, experimentally challenging. 

In this section, we aim to elucidate and model how heterogeneous pressure builds up inside the onion composite during rupture with a focus on the first-layer compression of soft onion tissue. 
The compression mechanism can be approximated with a flat-punch indentation model, in which the narrow blade exhibits a vertical load along the $\hat{z}$ direction on the onion surface, treated as a semi-infinite flat body \cite{johnson1987contact}. The size of the resulting stress zone, i.e., the compressed region in the blade-onion contact, is denoted as $L^*$, with $L^*\approx 0.1L_o\approx 0.3$ mm, where $L_o$ is the thickness of a single onion layer. Since $L^*$ is much smaller than both the onion's radius of curvature and the blade length ($L_b \simeq$ 100 mm), the problem reduces effectively to a two-dimensional $xz$-plane strain condition. 

The key difference from a classic flat punch problem lies in the onion layer’s sandwiched structure, where a tough epidermis (onion skin) encases a much softer mesophyll interior. Upon compression, a stress zone develops within the meosphyll, where most droplets generate from. The epidermis, however, would not fracture until the applied stress reaches its critical stress $\sigma_c$ for skin. At that moment, the mesophyll sustains high pressure that exceeds its own fracture stress, which in turn is relieved by rupture. The upper panel schematics in Fig.~\ref{o3}A illustrate this process, with the corresponding visualization in the lower panel. From the moment of contact $t_{\mathrm{contact}}$, the onion skin undergoes significant displacement, stretching the epidermis, and storing energy into the underlying mesophyll. Just before rupture, the blade slows down as it deforms the skin, followed by a rapid breakthrough once the epidermis fails. This slow-down and spring-forward motion of the blade tip is evident in Supplementary Fig. 5 and in SI Movie S3. The onion layer is also observed to immediately rebound back after rupture, signature of elastic energy release. Microscope images include in the upper panel insets, demonstrating differences between mesophyll and epidermis. The epidermis exhibits algined, elongated cell structures, in contrast to the more isotropic mesophyll. It is crucial to note that the skin here refers strictly to the thin epidermis of each layer, not the brown outermost dry layer that is commonly removed before kitchen cutting. 

Instron testing with a cylindrical flat punch reveals that the peak force required to fracture skin-on onion samples is typically two to three times higher than that for skin-off samples, as shown in Fig.~\ref{o3}B. This indicates that more energy is needed to fracture the skin-on samples, confirming the mechanical reinforcement provided by the thin outer epidermis layer. We further note that the onset of fracture on the skin-on samples is clearly aligned with the peak force, whereas the skin-off samples crack intermittently as the force increases in discrete steps. By examing the relatively linear portion of the $F-\delta$ curve, the elastic modulus $E$ can be estimated as $E=F(1-\nu^2)/(2a\delta)$ \cite{sneddon1965relation}, where $a$ is the indentor/blade size and $\nu$ the Poisson ratio. This further demonstrates that skin-on samples exhibit a significantly higher elastic modulus approximately 2.2 $\pm$ 0.36 MPa compared to 1.4 $\pm$ 0.3 MPa for skin-off samples, validating the role of the skin/epidermis in enhancing material stiffness. The measured values are consistent with the previously reported range of 2.0--5.4 MPa (skin-on samples) \cite{jafari2016determination}. The evidence supports that, at the moment of epidermal rupture and pressure release, the underlying mesophyll interior is sustaining internal stresses exceeding its fracture threshold. 

Using digital image correlation (DIC), we further extract strain distributions on the $xz$ cross section to visualize stress concentrations beneath and around the blade tip. A horizontal profile across the $\epsilon_{z}$ field (Fig.~\ref{o3}C) shows a localized stress zone approximately 0.3$L_o$ in width. Time-lapse data show the progressive growth of $\epsilon_z$ up to the moment of fracture at $t=$1.75 ms.
Fig.~\ref{o3}D, E, \& F present the time-lapse maps of Lagrangian strains $\epsilon_z$, $\epsilon_{xz}$, and $\epsilon_x$ on the $xz$ plane, respectively. In Fig.~\ref{o3}D, the regions with positive $\hat{z}$ tensile strain appear directly above the negatively compressive strain zone in $t=$5.00--6.75 ms, indicating that central compression due to the blade induces stretching in the surrounding regions. Fig.~\ref{o3}E shows the development of two oppositely signed shear strain regions on either side of the cutting direction as expected. Fig.~\ref{o3}F shows that red regions in $\epsilon_x$ show horizontal stretching within the stress zone due to compression, which suggests a Poisson ratio of $\nu=-\epsilon_x/\epsilon_z=$ 0.3--0.4, consistent with previously reported 0.26--0.45 \cite{jafari2016determination}. The complete  DIC movie is available in SI Movie S4. 
An unexpected region of negative $\epsilon_x$ appears directly beneath the blade, overlapping with the negative $\epsilon_z$ region. This indicates that the material directly below the blade tip experiences collapse in both vertical and horizontal directions, i.e., a signature of prematurely ruptured tissues prior to skin fracture. It causes some errors that limit the rigor in DIC analysis. For details on the stress zone analysis via DIC, see Supplementary Note 3 and Supplementary Fig. 6. 

%At $t=$1.50 ms, there exists a tensile region before compression, likely due to the blade tip first compressing on an $xz$ plane directly behind the observed plane. 

\subsection{\label{sec:level2} Connecting the initial droplet velocity $V_0$ to blade width $r_\mathrm{b}$}
In this section, we develop a simplified model to relate the blade tip width $r_\mathrm{b}$, strain $\epsilon$, and initial droplet velocity $V_0$. At the moment of rupture, the critical fracture stress and force on the skin can be written as $\sigma_{c,\mathrm{skin}}\propto F_c/r_\mathrm{b}L_c$ assuming the stress-strain linearity. This indicates $F_c \propto \sigma_{c,\mathrm{skin}}r_\mathrm{b}L_c$ as illustrated in Fig. ~\ref{o4}A, where $L_c$ is the blade-tissue contact length in the $\hat{y}$ direction. However, Instron data in Fig.~\ref{o4}B show a power law of $F_c\sim r_\mathrm{b}^{l}$ with $l\simeq$ 0.82. This deviation suggests that the force-width relationship is more complex than the linear model predicts. In particular, the pressurization within the stress zone likely exceeds the mesophyll's fracture stress, but the pressure distribution is not known \textit{a priori} and differs from the material critical fracture stress $\sigma_{c,\mathrm{skin}}$ at the blade tip. Such pressure distribution is highly dependent on $r_\mathrm{b}$. This distribution is highly dependent on $r_\mathrm{b}$ and directly determines the initial velocity of ejected droplets. Due to the difficulty in resolving the full pressure profile, we approximate the average strain in the hemispherical stress zone as $\epsilon_c\propto \delta_c/L^*$, where $\delta_c$ is the critical indentation depth at rupture, which can be extracted from experiments (see extraction details in Supplementary Note 3 and Supplementary Fig. 6), and $L^*$ is the characteristic length of the stress zone.

The relationship between blade tip width $r_\mathrm{b}$ and critical indentation depth $\delta_c$ is then extracted empirically. As expected, the linearity assumption does not hold for plastic deformation of cellular materials at fracture. Fig.~\ref{o4}C shows the force-displacement $F$-$\delta$ relationships from quasi-static ($\sim10^{-5}$ m/s) to typical cutting speeds ($10^{-1}$--$10^{0}$ m/s), using both needle (N) and blade (B) indentors. The data normalized by values at the moment of fracture show a power-law relationship of $\bar{F} \sim \bar{\delta}^m$, where overbars denote normalized variables and the exponent $m$ ranges from 1 to 3/2. This scaling leads to $\delta_c\sim F_c^{1/m}\sim r_\mathrm{b}^{l/m}$ with $l/m\simeq$ 0.53 -- 0.82. Fig.~\ref{o4}D presents the measured indentation depth $\delta_c$ as a function of $r_\mathrm{b}$, confirming the deeper indentation before fracture for larger balde widths. The parametric relationship is validated in Fig.~\ref{o4}E, showing strong agreement with the experimental data. 
 
 %Classical flat punch solution for an elastic semi-infinite flat domain, $\delta\sim \ln[(x+r_\mathrm{b})/r_\mathrm{b}]^2$ \cite{johnson1987contact}, fails at smaller length scale as shown in red in Fig.~\ref{o4}D, expected due to the bi-layer and increasing non-linearity at fracture. 

Given $\epsilon_c\sim r_\mathrm{b}^{l/m}$, the non-linearity between strain $\epsilon$ and cellular pressure $P_{\mathrm{cell}}$ is further addressed for soft cellular materials. A superlinear relationship between the two, as in \cite{zhu2003mechanics} for vegetable tissues, suggests a power law of $\epsilon^n\sim P_{\mathrm{cell}}$ with $n \simeq $ 1.9--2.3. Combining these relations, we express the pressure at fracture as $P_{\mathrm{cell}}\sim r_\mathrm{b}^{nl/m}$, and the corresponding initial droplet velocity as $V_\mathrm{d,0}\sim r_\mathrm{b}^{nl/2m}$ with $nl/2m\simeq$ 0.52 -- 0.94. By singling out the very early-stage data (i.e., droplets ejected within the first 0.5 ms), we extracted experimental values that align more closely with the lower end of this predicted range, as shown in Fig.~\ref{o4}F. 

\subsection{\label{sec:level2} Modeling with membrane on a spring foundation}

\begin{figure*}[h!tbp]
    \begin{center}
    \centering
  \vspace{-10pt}
    \includegraphics[width=1\textwidth]{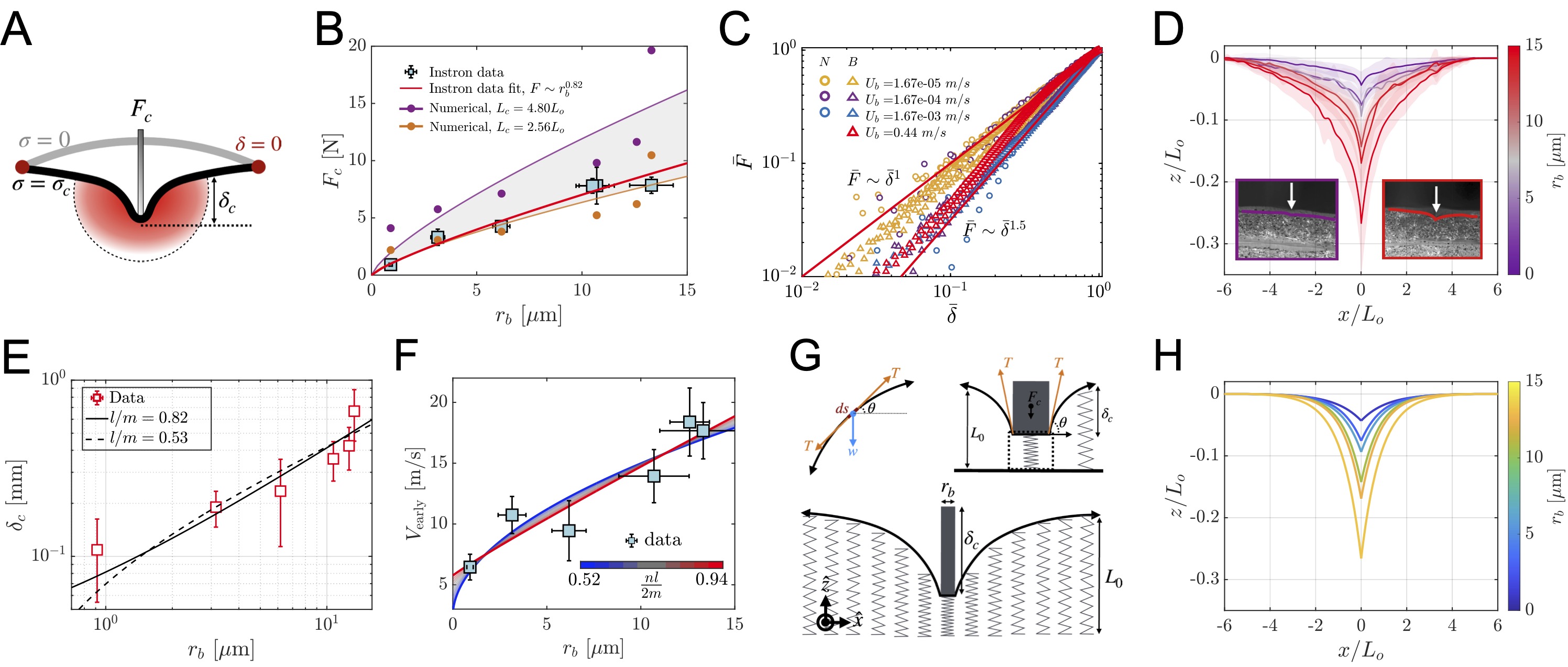}
     \caption{\textbf{Correlating blade sharpness to ejection speed.} \textbf{A.} Schematics here showcases the compression of the first layer, defining at the moment of fracture, critical stress $\sigma_c$, and critical displacement on the surface $\delta_c$. \textbf{B.} Instron measurements of critical force at fracture $F_c$ with range of predictions from fitting numerical solutions of critical indentation depth, for empirically measured contact length $L_o$. \textbf{C.} Normalized force versus displacement curves for onion before fracture, indented at different speed and indentor type: N for needle and B for blade, normalized at fracture point. Curves of exponent $m=$ 1 and 3/2 are given. \textbf{D.} Surface profile displacement $\delta$ at the moment of fracture for different blade sharpness, normalized by onion layer thickness. Shaded regions are the standard errors of the displacement data. Inset showcases the images of surface displacement for the sharpest (left) and the bluntest (right) sample at the point of fracture. \textbf{E.} Max displacement at the point of fracture measured empirically in relation to $r_\mathrm{b}$. Fits of supposed exponent $l/m=$ 2/3 and 1 are given. \textbf{F.} Experimental data with fitting of early-stage ejection velocity for the first layer in relation to $r_\mathrm{b}$. Color coding applied for different exponents, theoretically $nl/2m$.  \textbf{G.} Schematics of indentation model for an elastic membrane overlaid on an elastic space. Inset shows the free-body force diagram and the membrane tension force balance. \textbf{H.} Indentation profiles solved from numerical solutions of the spring foundation model, with empirical boundary conditions. %\red{1) Let's reduce the empty space between the graphs. 2) The order of figure is not straightforward. 4) Should we put 4G earlier?  5) In 4G, "Fit" means the power of 0.82? 6) Would you like to merge 4C and the inset of 4G? \blue{I greyed them out but I do wish to keep the lines though to match with the error bar.} 7) Fig. ~\ref{o4}E, please put two lines only and add a shade in between. } 
     }
    \label{o4}
    \end{center}
\end{figure*}

To further understand the mechanical response, we model the onion skin as an inextensible membrane within a 2D slice, assuming zero thickness compared to the onion layer scale $L_o$. The soft layer beneath is modeled as a spring foundation. This model considers a 2D cross-section of the onion under planar loading, where the membrane undergoes tension while the underneath spring foundation experiences compression (see Fig.~\ref{o4}G). 

For an infinitesimal piece of the membrane has its arc length as $ds$, the vertical force equilibrium on the membrane can be written as \cite{synge2011principles}:

\begin{equation} \label{dist}
\begin{split}
\frac{d}{ds}\left(\frac{dz}{dx}\right)&=-\frac{w}{H} \,,
\end{split}
\end{equation}
where $w$ is the distributed load per unit length, $z(t)$ is the vertical location of the membrane, and $H$ is the horizontal component of membrane tension $T$. For the force balance on the membrane, $H$ is assumed to be a constant for each case \cite{synge2011principles}. 
Here, the vertical load $w$ acting on an infinitesimal piece of the membrane can be written based on the elastic foundation as $w=E\,{(L_o-z(x))} / {L_o}$, representing the supporting force from the compressed mesophyll layer, where $L_o-z(x)$ denotes the local compression, and $E$ is Young's modulus of the mesophyll tissue approximately 1.4 MPa. By placing the origin to the peak indentation point at fracture and nondimensionalizing length scales by $L_o$, the governing equation becomes 

%Units may initially seem at odd between $E$ and a spring constant. However, $w$ here is force per length; thus the spring constant $K$ carries a unit of $\mathrm{N/m}^2$. We can imagine the string and the spring base here has certain out of plane dimension size $\delta{y}$ in $\hat{y}$. We can then immediately see that:

%\begin{equation} \label{dist}
%\begin{split}
%E dx \delta y \frac{\delta(x)}{L_0}=K dx \delta(x).
%\end{split}
%\end{equation}

%\delta (x)=L_0-z(x)$ is the compression amount across the profile. To simplify analysis and ensure $E=K$, we take here the out-of-plane thickness for the model as $\delta y=L_0$, onion layer thickness of 3 mm, the depth at which the indentation is no longer felt.  

\begin{equation} \label{dist}
\begin{split}
\frac{\tilde{z}^{''}}{\sqrt{1+\tilde{z}^{'2}}}=-\gamma ( \tilde{\delta_c}-\tilde{z}) \,,
\end{split}
\end{equation}
where $\gamma={EL_o}/{H}$ is a non-dimensional parameter comparing the vertical spring force to the horizontal component of membrane tension, controlling the curvature profile. The normalized vertical profile is $\tilde{z}=(z-(L_o-\delta_c))/L_o$ with origin shifted to the critical fracture point and the normalized critical indentation depth is $\tilde{\delta}_c=\delta_c/L_o$. The normalized critical indentation depths $\tilde{\delta}_c$ are experimentally extracted from Fig.~\ref{o4}E. This second order nonlinear ODE is solved numerically using a shooting method starting with a initial condition $\tilde{z}(0)=0$ and satisfying a boundary condition $\tilde{z}^{'}(\tilde{x}_{\mathrm{end}})=0$ at the far end of the domain. With the governing equation and boundary conditions, we solve for the vertical displacement of the membrane and the unknown $\gamma$ is determined from the relation $\tilde{z}(\tilde{x}_{\mathrm{end}}) - \tilde{z}(0)= \tilde{\delta}_c$. From Fig.~\ref{o4}D, the profile reaches a plateau with zero slope at around $x/L_o\simeq$5 for all blade widths, which we take as the far-end boundary location $\tilde{x}_{\mathrm{end}}$. The resulting surface profiles from numerical calculations are plotted in the Fig.~\ref{o4}H.%\red{Can you check the previous two sentences? } \blue{yes, just did. Please check.}

The resulting force, $H$, is determined from the solved parameter $\gamma$. The critical fracture force, $F_c$, is then calculated from a free-body analysis, as illustrated in the schematic inset of Fig.~\ref{o4}G. 

\begin{equation} \label{dist}
\begin{split}
F_c=EL_c \left( a\frac{\delta_c}{L_o}+L_o \frac{2z'(0)}{\gamma} \right) \,,
\end{split}
\end{equation}
where $L_c$ is the blade-onion contact length ($\approx$9.2 $\pm$ 2.8 mm from experiments, i.e., around 3 -- 5$L_o$). The first term represents the upward force from the spring foundation and the second term is the tensile force from the membrane, $T\mathrm{sin}(\theta)$. The resulting force is shown as a shaded grey region in Fig.~\ref{o4}B bounded by two fitted lines (purple and yellow), corresponding to the upper (4.80$L_o$) and lower (2.56$L_o$) limits of $L_c$. Both fits are $R^2\approx 0.72$, 
%\red{Can you report $R^2$ separately for the lower and upper limits?} \blue{I do not understand. they are the same R2?} \red{There are two fits and two data sets; one for the upper limit and the other for the lower limit. Should we report $R^2$ for each fit? } \blue{they are the same simulated force, just times a different Lc value, so the fit has exactly the same value for R2.}
and the independently measured fracture forces from Instron tests fall within the predicted range, validating the above model. 

% \red{I am a bit confused about the goal of this section. Do you want to get the relation of $F_c  \sim r_\mathrm{b}^{0.82}$? Then, how is the right-hand-side term in Eq. (5) related to $r_\mathrm{b}$? Or by solving the ODE, we get curves close to our experimental results? } \blue{Hi Sunny, I have remade this section a bit. Please let me know if eveyrthing works. I checked all the data. everything is correct. I indeed only have 6 expts for 4B, one of them  I did not do the Instron fracture test. If need be, I can try to add it with new expt. Also, for 4F, I checked, I used a new set of zoomed in expt at cutting height of 3cm and one of the blades I snapped in half to try to do the cross section SEM, so I was not able to get new data for that particular blade during the test, thus the 6 data point there. But I can assure there are no data points neglected here. I previously used the first 30-40 drops from each video, but that seems incorrect for $V_{early}$, so I ended up counting all droplets before 0.5 ms in each video, thus the data look slightly different here, but I believe that logic of measurements is more consisten.}
%Literature has reported values for fracture strain of epidermal tissues up to 36-45$\%$ \cite{zamil2015mechanical}.

\section{\label{sec:level1}Discussion}

In summary, we investigated the mechanism of droplet generation during onion cutting under different conditions, primarily cutting speed and blade sharpness. Using our custom-developed particle visualization and tracking techniques, we revealed that oninon cutting with unsharpened blades generates liquid droplets through a two-step process: an initial outburst driven by internal pressurization, followed by a slower fragmentation of ligaments in air. 
The emerging onion droplets are concentrated within a nearly vertical plane, tilted approximately 8$\degree$ from the blade surface. Exit velocities decay over time, reflecting the increased resistance encountered by fluid originating from deeper onion layers. Both blade cutting speed $U_\mathrm{b}$ and blade tip width $r_\mathrm{b}$ effectively increase the number and average energy of ejected droplets. 

Focusing on droplet ejection from the first onion layer, which exhibits the higher initial velocities, we developed high-speed DIC procedures to extract the temporal evolution of the cross-sectional strain maps during onion cutting. These maps reveal the formation of a localized stress zone immediately beneath the blade tip. The strain maps also suggest a shielding effect made by the tough onion epidermis that allows the softer mesophyll beneath to accumulate excessive stress before rupture. This effect enables blunter blades to indent deeper, causing greater pressurization before fracture. A scaling analysis based on this mechanism explains well the observation of the early-stage droplet velocity $V_\mathrm{d}$ increasing with blade tip width $r_\mathrm{b}$, following the relationship $V_\mathrm{d,0}\sim r_\mathrm{b}^{nl/2m}$ with $nl/2m=$0.52 -- 0.94.
Finally, we developed a simplified model representing the onion composite layer as an inextensible membrane supported by a spring foundation. Numerical solutions with empirical boundary conditions predict the fracture forces across different blade widths $r_\mathrm{b}$, with results in great agreement with independently measured Instron data. 

We also evaluated common strategies for reducing droplet ejection using blunter blades with $r_\mathrm{b}\approx$ 10--12 $\mu$m, along with the PTV and indentation profile techniques in order to assess the effects of cutting orientation and temperature. However, these effects were difficult to isolate due to substantial sample variations. Both cutting orientations produced similar droplet velocities, normalized by the mean speed from pole-to-pole condition as $\overline{V_\mathrm{d}}=V_\mathrm{d}/ \langle V_\mathrm{d,P} \rangle$. %\red{Is it normalized velocity by what? Did we define $V_\mathrm{d,P}$? I don't see it earlier. }. \blue{it is normalized by the mean speed from pole condition $\langle V_\mathrm{d,P} \rangle$} 
The median $\overline{V_\mathrm{d}}$ from \textit{equator} cuts was slightly higher than that from \textit{pole-to-pole} cuts. The critical indentation profiles were almost identical. These findings are summarized in Supplementary Fig.~7A and B. 
Previous studies have shown that indenting plant tissues along the longitudinal cell direction (\textit{pole} direction in onion epidermis) requires lower fracture energy than transverse directions \cite{vincent1990fracture,bidhendi2023cell}, which may explain the marginal increase in droplet speed from \textit{equator} cuts. %\red{Was the previous test done with onion?} \blue{various plants, the second study includes some onion samples} 
However, this difference was not statistically significant ($p$=0.6 for $\overline{V_\mathrm{d}}$ from Kolmogorov–Smirnov test). 

A common perception is that chilling onions prior to cutting reduces enzymatic activity and associated irritation \cite{Epicurious}. To investigate whether cooling also alters the mechanical response during cutting, we refrigerated onions at 1 $\degree$C for 12 hours and compared their cuts with room-temperature samples. Similarly, Statistical analysis of droplet velocity distributions showed no significant difference between the two conditions with $p$=0.4. The corresponding fracture profiles were nearly identical, as shown in Supplementary Fig.~7C and D. However, we observed a noticeably larger volume of droplet ejection from the chilled samples. Previous studies have reported increased elasticity and brittleness of plant tissues at lower temperatures  \cite{vincent1990fracture,treitel1946elasticity}, which may have expanded the fracture region, even though the critical epidermal fracture threshold remained unchanged. Further controlled experiments are needed to rigorously evaluate the mechanical effects of temperature on onion cutting and droplet generation. 

Food-borne pathogens in kitchens are known to spread through droplet splashes during washing and direct surface contact \cite{lai2022handling}. A study of commercial kitchens reported that 33 -- 100$\%$ of cutting and washing regions were contaminated with \textit{Campylobacter} \cite{lai2022handling}, a common cause of diarrhea. The current work reveals an alternate route for bacterial agents to spread via fracture from raw food cutting: inside fragmented food droplets. This is supported by previous work showing contamination levels decrease with distance from cutting boards during chicken chopping \cite{lai2023evaluation}. Ejected droplets can come into direct contact with contaminated blades or carry surface-borne pathogens as they leave the food surface. The corresponding Stokes numbers of the droplets can be calculated as $\mathrm{St}=\tau U_\mathrm{b}/L$ with the characteristic time for droplets defined as $\tau=\rho_\mathrm{d}D_\mathrm{d}^2/18 \mu_\mathrm{a}$, on the order of 10$^{-2}$ -- 10$^{0}$; $L$ is taken as the size of the onion and $\mu_\mathrm{a}$ is the air dynamic viscosity at 1.8$\times$10$^{-5}$ Pa$\cdot$s. While heavier droplets with larger sizes follow its inertial pathways, lighter droplets is readily suspended and can be transported by ambient air currents, thereby posing a potential risk for airborne transmission. 

The current findings reveal the importance of blade sharpening during raw food cutting practices, as sharper blades reduce not only the number of droplets but also their speed and kinetic energy. This is particularly relevant for fruits and vegetables, which can carry food-borne pathogens such as \textit{Salmonella} \cite{berger2010fresh}, and typically possess tough epidermal layers allowing excessive internal compression before fracture as described in this study.

In summary, motivated by the mechanics underlying onion tear induction and pathogen-laden droplet ejection during cutting, we developed customized PTV and DIC experimental procedures to visualize, quantify, and link droplet generation to cellular compression in onions. These observations were supported by theoretical models that accurately capture independently measured fracture forces. Our findings demonstrate that blunter blades increase both the speed and number of ejected droplets, providing experimental validation for the widely held belief that sharpening knives reduces onion-induced tearing. Beyond comfort, this practice also plays a critical role in minimizing the spread of airborne pathogens in kitchens, particularly when cutting vegetables with tough outer layers capable of storing significant elastic energy prior to rupture.

\section{\label{sec:level1} Methods}
\subsection{Sample preparation and characterizations}
Steel blades were purchased from Jsumo Co. and cut to dimensions of 5 mm (width) by 200 mm (length). To sharpen the blades, we used 3M aluminum oxide abrasive films with 0.3 -- 30 $\mu$m grits. The sharpness is then tuned by first holding the blade vertical and translating it on top of the film horizontally. The blade was rotated to match the wedge angle and polished by moving it along the wedge surface. The sharpness of the blade was characterized using scanning electron microscopy (Zeiss LEO 1550 FESEM) at the Cornell Center for Materials Science (CCMR). The blade was attached to the measurement stage with the cutting edge oriented toward the electron beam. 7--10 measurements were taken for each blade across the sample.

Onions were always freshly sourced from local vendors and experimented within two weeks. The species \textit{Allium cepa L.} was chosen for consistency. The outermost dry layer (brown skin) was removed, which is typically not used in culinary applications. The onion was then separated into either two halves for regular PTV measurements or four quarters for high-speed DIC experiments. 

For PIV tests on speed variations with 3 conditions, we conducted 3 spatially separated cuts on the same onion half to reduce sample variations. Adhesive tape was used to secure the onion and to cover previous cut openings to limit fluid ejection to the active cutting zone. For blade sharpness variations, tests were conducted on different onions to collect averaged results across samples. The overall ejection speed statistics show in Fig.~\ref{o2_2} were obtained PTV measurements. For early-stage ejection speeds before 0.5 ms in Fig.~\ref{o4}F, we ddressed potential early-time PTV errors caused by dense droplet ensembles from videos at a higher frame rate, 20,000 fps. The velocities of all droplets within the first 0.5 ms (first 10 frames) were then manually measured.

\subsection{Error estimation with $yz$ projection}
The angular distribution of droplet ejection $\theta$ shows that 80$\%$ of the ejection data fall within a 20\% error margin for velocity $V_\mathrm{d}$ when using the $yz$ projection. This means that for slower droplets below 8 m/s, which may have errors above 30$\%$ displacing the worst-case-scenario 1-mm depth of field at 8000 fps, most of the error is still limited to 20$\%$. These results validate the use of the $yz$ projection for characterizing the statistical properties of the droplet ensemble. 

\subsection{Segmentation analysis from ejection}
High-speed video recordings were subjected to segmentation analysis to differentiate between external structures (onion and blade), ligaments, and droplets. Classification was based on geometric features such as size, circularity, and eccentricity, along with spatial filters. Additional boundary-based recognition techniques were employed to identify droplets exiting the frame, enabling accurate measurement of cumulative droplet count and volume.

 %The samples are grown from local farm of K.S. Datthyn, based in Sodus, NY. 
\subsection{\label{sec:level2} High-speed digital image correlation (DIC)
}

High-speed digital image correlation was conducted with the following steps: speckle patterning, high-speed videography, and DIC analysis. A quarter section of an onion was used for each trial. To create a suitable speckle pattern, the exposed cross-section was coated with black spray paint (Rust-Oleum commercial brand) applied from a horizontal distance of 30–40 cm, followed by 30 minutes of drying.
High-speed imaging was performed using Photron FASTCAM Nova and SA-Z series cameras, operating at frame rates ranging from 8,000 to 20,000 fps. The resulting videos were segmented into individual frames and analyzed using Ncorr, an open-source 2D DIC program in MATLAB. 
The reference frame was selected as the last frame prior to blade–onion contact. Subset radii for correlation were set between 10–15 pixels, corresponding to approximately 200–400 $\mu$m depending on resolution, and were selected to match the scale of onion cells. Seed propagation and step analysis were enabled to accommodate large deformations. For strain calculation, a strain radius of 7–15 pixels was empirically chosen to minimize error, as recommended in the program manual. The resulting strain maps were exported for further analysis and visualization. 

\begin{acknowledgments}

\end{acknowledgments}
This work made use of the rheometer (TA Instruments DHR3 Rheometer) and electron microscopy facilities (Zeiss LEO 1550 FESEM (Keck SEM))) of the Cornell Center for
Materials Research (CCMR) with support from the National Science Foundation
Materials Research Science and Engineering Centers (MRSEC) program (DMR1719875). 

The authors would like to thank Ms. Marielle Mullon and Ms. Meredith Brafman for their valuable assistance in setting up and performing some of the experiments and analysis. 

%\section*{Data availability}
%The data used in this study are deposited in the public repository.

\section*{Author contributions}
S.J. conceived the idea. Z.W., W.W. and A.H. performed experiments. Z.W. and A.H. analyzed the data. Z.W. wrote the original draft. C.H. helped develop the bi-layer mechanical model. All authors revised the manuscript and contributed to the final version. 

\section*{Competing interests}
The authors declare no competing interests.

% \bibliography{ref}% Produces the bibliography via BibTeX.

\begin{thebibliography}{49}%
\makeatletter
\providecommand \@ifxundefined [1]{%
 \@ifx{#1\undefined}
}%
\providecommand \@ifnum [1]{%
 \ifnum #1\expandafter \@firstoftwo
 \else \expandafter \@secondoftwo
 \fi
}%
\providecommand \@ifx [1]{%
 \ifx #1\expandafter \@firstoftwo
 \else \expandafter \@secondoftwo
 \fi
}%
\providecommand \natexlab [1]{#1}%
\providecommand \enquote  [1]{``#1''}%
\providecommand \bibnamefont  [1]{#1}%
\providecommand \bibfnamefont [1]{#1}%
\providecommand \citenamefont [1]{#1}%
\providecommand \href@noop [0]{\@secondoftwo}%
\providecommand \href [0]{\begingroup \@sanitize@url \@href}%
\providecommand \@href[1]{\@@startlink{#1}\@@href}%
\providecommand \@@href[1]{\endgroup#1\@@endlink}%
\providecommand \@sanitize@url [0]{\catcode `\\12\catcode `\$12\catcode
  `\&12\catcode `\#12\catcode `\^12\catcode `\_12\catcode `\%12\relax}%
\providecommand \@@startlink[1]{}%
\providecommand \@@endlink[0]{}%
\providecommand \url  [0]{\begingroup\@sanitize@url \@url }%
\providecommand \@url [1]{\endgroup\@href {#1}{\urlprefix }}%
\providecommand \urlprefix  [0]{URL }%
\providecommand \Eprint [0]{\href }%
\providecommand \doibase [0]{https://doi.org/}%
\providecommand \selectlanguage [0]{\@gobble}%
\providecommand \bibinfo  [0]{\@secondoftwo}%
\providecommand \bibfield  [0]{\@secondoftwo}%
\providecommand \translation [1]{[#1]}%
\providecommand \BibitemOpen [0]{}%
\providecommand \bibitemStop [0]{}%
\providecommand \bibitemNoStop [0]{.\EOS\space}%
\providecommand \EOS [0]{\spacefactor3000\relax}%
\providecommand \BibitemShut  [1]{\csname bibitem#1\endcsname}%
\let\auto@bib@innerbib\@empty
%</preamble>
\bibitem [{\citenamefont {Pareek}\ \emph {et~al.}(2017)\citenamefont {Pareek},
  \citenamefont {Sagar}, \citenamefont {Sharma},\ and\ \citenamefont
  {Kumar}}]{pareek2017onion}%
  \BibitemOpen
  \bibfield  {author} {\bibinfo {author} {\bibfnamefont {S.}~\bibnamefont
  {Pareek}}, \bibinfo {author} {\bibfnamefont {N.~A.}\ \bibnamefont {Sagar}},
  \bibinfo {author} {\bibfnamefont {S.}~\bibnamefont {Sharma}},\ and\ \bibinfo
  {author} {\bibfnamefont {V.}~\bibnamefont {Kumar}},\ }\bibfield  {title}
  {\bibinfo {title} {Onion (allium cepa l.)},\ }\href@noop {} {\bibfield
  {journal} {\bibinfo  {journal} {Fruit and Vegetable Phytochemicals: Chemistry
  and Human Health, 2nd Edition}\ }\textbf {\bibinfo {volume} {2}},\ \bibinfo
  {pages} {1145} (\bibinfo {year} {2017})}\BibitemShut {NoStop}%
\bibitem [{\citenamefont {Lanzotti}(2006)}]{lanzotti2006analysis}%
  \BibitemOpen
  \bibfield  {author} {\bibinfo {author} {\bibfnamefont {V.}~\bibnamefont
  {Lanzotti}},\ }\bibfield  {title} {\bibinfo {title} {The analysis of onion
  and garlic},\ }\href@noop {} {\bibfield  {journal} {\bibinfo  {journal}
  {Journal of chromatography A}\ }\textbf {\bibinfo {volume} {1112}},\ \bibinfo
  {pages} {3} (\bibinfo {year} {2006})}\BibitemShut {NoStop}%
\bibitem [{nms()}]{nmsuHistoryMexico}%
  \BibitemOpen
  \href@noop {} {\bibinfo {title} {New mexico state university onion breeding
  program}},\ \bibinfo {howpublished}
  {https://onion.nmsu.edu/index.html}\BibitemShut {NoStop}%
\bibitem [{\citenamefont {Block}(2010)}]{block2010garlic}%
  \BibitemOpen
  \bibfield  {author} {\bibinfo {author} {\bibfnamefont {E.}~\bibnamefont
  {Block}},\ }\href@noop {} {\emph {\bibinfo {title} {Garlic and other
  alliums}}}\ (\bibinfo  {publisher} {RSC publishing, Cambridge},\ \bibinfo
  {year} {2010})\BibitemShut {NoStop}%
\bibitem [{\citenamefont {Kato}\ \emph {et~al.}(2016)\citenamefont {Kato},
  \citenamefont {Masamura}, \citenamefont {Shono}, \citenamefont {Okamoto},
  \citenamefont {Abe},\ and\ \citenamefont {Imai}}]{kato2016production}%
  \BibitemOpen
  \bibfield  {author} {\bibinfo {author} {\bibfnamefont {M.}~\bibnamefont
  {Kato}}, \bibinfo {author} {\bibfnamefont {N.}~\bibnamefont {Masamura}},
  \bibinfo {author} {\bibfnamefont {J.}~\bibnamefont {Shono}}, \bibinfo
  {author} {\bibfnamefont {D.}~\bibnamefont {Okamoto}}, \bibinfo {author}
  {\bibfnamefont {T.}~\bibnamefont {Abe}},\ and\ \bibinfo {author}
  {\bibfnamefont {S.}~\bibnamefont {Imai}},\ }\bibfield  {title} {\bibinfo
  {title} {Production and characterization of tearless and non-pungent onion},\
  }\href@noop {} {\bibfield  {journal} {\bibinfo  {journal} {Scientific
  Reports}\ }\textbf {\bibinfo {volume} {6}},\ \bibinfo {pages} {23779}
  (\bibinfo {year} {2016})}\BibitemShut {NoStop}%
\bibitem [{\citenamefont {Spare}\ \emph {et~al.}(1963)\citenamefont {Spare},
  \citenamefont {Virtanen}, \citenamefont {Munch-Petersen}, \citenamefont
  {Kvande},\ and\ \citenamefont {Meisingseth}}]{spare1963lachrymatory}%
  \BibitemOpen
  \bibfield  {author} {\bibinfo {author} {\bibfnamefont {C.}~\bibnamefont
  {Spare}}, \bibinfo {author} {\bibfnamefont {A.~I.}\ \bibnamefont {Virtanen}},
  \bibinfo {author} {\bibfnamefont {J.}~\bibnamefont {Munch-Petersen}},
  \bibinfo {author} {\bibfnamefont {P.}~\bibnamefont {Kvande}},\ and\ \bibinfo
  {author} {\bibfnamefont {E.}~\bibnamefont {Meisingseth}},\ }\bibfield
  {title} {\bibinfo {title} {On the lachrymatory factor in onion (allium cepa)
  vapours and its precursor},\ }\href@noop {} {\bibfield  {journal} {\bibinfo
  {journal} {Acta Chem. Scand}\ }\textbf {\bibinfo {volume} {17}},\ \bibinfo
  {pages} {641} (\bibinfo {year} {1963})}\BibitemShut {NoStop}%
\bibitem [{\citenamefont {Kohman}(1947)}]{kohman1947chemical}%
  \BibitemOpen
  \bibfield  {author} {\bibinfo {author} {\bibfnamefont {E.~F.}\ \bibnamefont
  {Kohman}},\ }\bibfield  {title} {\bibinfo {title} {The chemical components of
  onion vapors responsible for wound-healing qualities},\ }\href@noop {}
  {\bibfield  {journal} {\bibinfo  {journal} {Science}\ }\textbf {\bibinfo
  {volume} {106}},\ \bibinfo {pages} {625} (\bibinfo {year}
  {1947})}\BibitemShut {NoStop}%
\bibitem [{\citenamefont {Style}\ \emph {et~al.}(2021)\citenamefont {Style},
  \citenamefont {Tutika}, \citenamefont {Kim},\ and\ \citenamefont
  {Bartlett}}]{style2021solid}%
  \BibitemOpen
  \bibfield  {author} {\bibinfo {author} {\bibfnamefont {R.~W.}\ \bibnamefont
  {Style}}, \bibinfo {author} {\bibfnamefont {R.}~\bibnamefont {Tutika}},
  \bibinfo {author} {\bibfnamefont {J.~Y.}\ \bibnamefont {Kim}},\ and\ \bibinfo
  {author} {\bibfnamefont {M.~D.}\ \bibnamefont {Bartlett}},\ }\bibfield
  {title} {\bibinfo {title} {Solid--liquid composites for soft multifunctional
  materials},\ }\href@noop {} {\bibfield  {journal} {\bibinfo  {journal}
  {Advanced Functional Materials}\ }\textbf {\bibinfo {volume} {31}},\ \bibinfo
  {pages} {2005804} (\bibinfo {year} {2021})}\BibitemShut {NoStop}%
\bibitem [{\citenamefont {Mehrabian}\ \emph {et~al.}(2016)\citenamefont
  {Mehrabian}, \citenamefont {Harting},\ and\ \citenamefont
  {Snoeijer}}]{mehrabian2016soft}%
  \BibitemOpen
  \bibfield  {author} {\bibinfo {author} {\bibfnamefont {H.}~\bibnamefont
  {Mehrabian}}, \bibinfo {author} {\bibfnamefont {J.}~\bibnamefont {Harting}},\
  and\ \bibinfo {author} {\bibfnamefont {J.~H.}\ \bibnamefont {Snoeijer}},\
  }\bibfield  {title} {\bibinfo {title} {Soft particles at a fluid interface},\
  }\href@noop {} {\bibfield  {journal} {\bibinfo  {journal} {Soft Matter}\
  }\textbf {\bibinfo {volume} {12}},\ \bibinfo {pages} {1062} (\bibinfo {year}
  {2016})}\BibitemShut {NoStop}%
\bibitem [{\citenamefont {Arnbjerg-Nielsen}\ \emph {et~al.}(2024)\citenamefont
  {Arnbjerg-Nielsen}, \citenamefont {Biviano},\ and\ \citenamefont
  {Jensen}}]{arnbjerg2024competition}%
  \BibitemOpen
  \bibfield  {author} {\bibinfo {author} {\bibfnamefont {S.~F.}\ \bibnamefont
  {Arnbjerg-Nielsen}}, \bibinfo {author} {\bibfnamefont {M.~D.}\ \bibnamefont
  {Biviano}},\ and\ \bibinfo {author} {\bibfnamefont {K.~H.}\ \bibnamefont
  {Jensen}},\ }\bibfield  {title} {\bibinfo {title} {Competition between
  slicing and buckling underlies the erratic nature of paper cuts},\
  }\href@noop {} {\bibfield  {journal} {\bibinfo  {journal} {Physical Review
  E}\ }\textbf {\bibinfo {volume} {110}},\ \bibinfo {pages} {025003} (\bibinfo
  {year} {2024})}\BibitemShut {NoStop}%
\bibitem [{\citenamefont {Plummer}\ \emph {et~al.}(2024)\citenamefont
  {Plummer}, \citenamefont {Adkins}, \citenamefont {Louf}, \citenamefont
  {Ko{\v{s}}mrlj},\ and\ \citenamefont {Datta}}]{plummer2024obstructed}%
  \BibitemOpen
  \bibfield  {author} {\bibinfo {author} {\bibfnamefont {A.}~\bibnamefont
  {Plummer}}, \bibinfo {author} {\bibfnamefont {C.}~\bibnamefont {Adkins}},
  \bibinfo {author} {\bibfnamefont {J.-F.}\ \bibnamefont {Louf}}, \bibinfo
  {author} {\bibfnamefont {A.}~\bibnamefont {Ko{\v{s}}mrlj}},\ and\ \bibinfo
  {author} {\bibfnamefont {S.~S.}\ \bibnamefont {Datta}},\ }\bibfield  {title}
  {\bibinfo {title} {Obstructed swelling and fracture of hydrogels},\
  }\href@noop {} {\bibfield  {journal} {\bibinfo  {journal} {Soft matter}\
  }\textbf {\bibinfo {volume} {20}},\ \bibinfo {pages} {1425} (\bibinfo {year}
  {2024})}\BibitemShut {NoStop}%
\bibitem [{\citenamefont {Lilin}\ \emph {et~al.}(2024)\citenamefont {Lilin},
  \citenamefont {Elkhoury}, \citenamefont {Peters},\ and\ \citenamefont
  {Bischofberger}}]{lilin2024fracture}%
  \BibitemOpen
  \bibfield  {author} {\bibinfo {author} {\bibfnamefont {P.}~\bibnamefont
  {Lilin}}, \bibinfo {author} {\bibfnamefont {J.~E.}\ \bibnamefont {Elkhoury}},
  \bibinfo {author} {\bibfnamefont {I.~R.}\ \bibnamefont {Peters}},\ and\
  \bibinfo {author} {\bibfnamefont {I.}~\bibnamefont {Bischofberger}},\
  }\bibfield  {title} {\bibinfo {title} {Fracture and relaxation in dense
  cornstarch suspensions},\ }\href@noop {} {\bibfield  {journal} {\bibinfo
  {journal} {PNAS nexus}\ }\textbf {\bibinfo {volume} {3}},\ \bibinfo {pages}
  {pgad451} (\bibinfo {year} {2024})}\BibitemShut {NoStop}%
\bibitem [{\citenamefont {Watson}\ \emph {et~al.}(2018)\citenamefont {Watson},
  \citenamefont {Stephen},\ and\ \citenamefont {Dickerson}}]{watson2018jet}%
  \BibitemOpen
  \bibfield  {author} {\bibinfo {author} {\bibfnamefont {D.~A.}\ \bibnamefont
  {Watson}}, \bibinfo {author} {\bibfnamefont {J.~L.}\ \bibnamefont
  {Stephen}},\ and\ \bibinfo {author} {\bibfnamefont {A.~K.}\ \bibnamefont
  {Dickerson}},\ }\bibfield  {title} {\bibinfo {title} {Jet amplification and
  cavity formation induced by penetrable fabrics in hydrophilic sphere entry},\
  }\href@noop {} {\bibfield  {journal} {\bibinfo  {journal} {Physics of
  Fluids}\ }\textbf {\bibinfo {volume} {30}},\ 
  \bibinfo {pages} {082109}
  (\bibinfo {year}
  {2018})}\BibitemShut {NoStop}%
\bibitem [{\citenamefont {Wang}\ \emph {et~al.}(2022)\citenamefont {Wang},
  \citenamefont {Yang}, \citenamefont {Zhang}, \citenamefont {Ke},
  \citenamefont {Zhang}, \citenamefont {Yang},\ and\ \citenamefont
  {Wu}}]{wang2022effects}%
  \BibitemOpen
  \bibfield  {author} {\bibinfo {author} {\bibfnamefont {B.}~\bibnamefont
  {Wang}}, \bibinfo {author} {\bibfnamefont {J.}~\bibnamefont {Yang}}, \bibinfo
  {author} {\bibfnamefont {J.}~\bibnamefont {Zhang}}, \bibinfo {author}
  {\bibfnamefont {Z.}~\bibnamefont {Ke}}, \bibinfo {author} {\bibfnamefont
  {H.}~\bibnamefont {Zhang}}, \bibinfo {author} {\bibfnamefont
  {Y.}~\bibnamefont {Yang}},\ and\ \bibinfo {author} {\bibfnamefont
  {J.}~\bibnamefont {Wu}},\ }\bibfield  {title} {\bibinfo {title} {Effects of
  tongue hair flexural deformation on viscous fluid transport by bees},\
  }\href@noop {} {\bibfield  {journal} {\bibinfo  {journal} {Bioinspiration \&
  Biomimetics}\ }\textbf {\bibinfo {volume} {18}},\ \bibinfo {pages} {016009}
  (\bibinfo {year} {2022})}\BibitemShut {NoStop}%
\bibitem [{\citenamefont {Herrera-Amaya}\ \emph {et~al.}(2021)\citenamefont
  {Herrera-Amaya}, \citenamefont {Seber}, \citenamefont {Murphy}, \citenamefont
  {Patry}, \citenamefont {Knowles}, \citenamefont {Bubel}, \citenamefont
  {Maas},\ and\ \citenamefont {Byron}}]{herrera2021spatiotemporal}%
  \BibitemOpen
  \bibfield  {author} {\bibinfo {author} {\bibfnamefont {A.}~\bibnamefont
  {Herrera-Amaya}}, \bibinfo {author} {\bibfnamefont {E.~K.}\ \bibnamefont
  {Seber}}, \bibinfo {author} {\bibfnamefont {D.~W.}\ \bibnamefont {Murphy}},
  \bibinfo {author} {\bibfnamefont {W.~L.}\ \bibnamefont {Patry}}, \bibinfo
  {author} {\bibfnamefont {T.~S.}\ \bibnamefont {Knowles}}, \bibinfo {author}
  {\bibfnamefont {M.~M.}\ \bibnamefont {Bubel}}, \bibinfo {author}
  {\bibfnamefont {A.~E.}\ \bibnamefont {Maas}},\ and\ \bibinfo {author}
  {\bibfnamefont {M.~L.}\ \bibnamefont {Byron}},\ }\bibfield  {title} {\bibinfo
  {title} {Spatiotemporal asymmetry in metachronal rowing at intermediate
  reynolds numbers},\ }\href@noop {} {\bibfield  {journal} {\bibinfo  {journal}
  {Integrative and comparative biology}\ }\textbf {\bibinfo {volume} {61}},\
  \bibinfo {pages} {1579} (\bibinfo {year} {2021})}\BibitemShut {NoStop}%
\bibitem [{\citenamefont {Smith}\ \emph {et~al.}(2018)\citenamefont {Smith},
  \citenamefont {Ebrahimi}, \citenamefont {Ghosh},\ and\ \citenamefont
  {Dickerson}}]{smith2018high}%
  \BibitemOpen
  \bibfield  {author} {\bibinfo {author} {\bibfnamefont {N.~M.}\ \bibnamefont
  {Smith}}, \bibinfo {author} {\bibfnamefont {H.}~\bibnamefont {Ebrahimi}},
  \bibinfo {author} {\bibfnamefont {R.}~\bibnamefont {Ghosh}},\ and\ \bibinfo
  {author} {\bibfnamefont {A.~K.}\ \bibnamefont {Dickerson}},\ }\bibfield
  {title} {\bibinfo {title} {High-speed microjets issue from bursting oil gland
  reservoirs of citrus fruit},\ }\href@noop {} {\bibfield  {journal} {\bibinfo
  {journal} {Proceedings of the National Academy of Sciences}\ }\textbf
  {\bibinfo {volume} {115}},\ \bibinfo {pages} {E5887} (\bibinfo {year}
  {2018})}\BibitemShut {NoStop}%
\bibitem [{\citenamefont {Box}\ \emph {et~al.}(2020)\citenamefont {Box},
  \citenamefont {Jacquemot}, \citenamefont {Adda-Bedia},\ and\ \citenamefont
  {Vella}}]{box2020cloaking}%
  \BibitemOpen
  \bibfield  {author} {\bibinfo {author} {\bibfnamefont {F.}~\bibnamefont
  {Box}}, \bibinfo {author} {\bibfnamefont {C.}~\bibnamefont {Jacquemot}},
  \bibinfo {author} {\bibfnamefont {M.}~\bibnamefont {Adda-Bedia}},\ and\
  \bibinfo {author} {\bibfnamefont {D.}~\bibnamefont {Vella}},\ }\bibfield
  {title} {\bibinfo {title} {Cloaking by coating: how effectively does a thin,
  stiff coating hide a soft substrate?},\ }\href@noop {} {\bibfield  {journal}
  {\bibinfo  {journal} {Soft Matter}\ }\textbf {\bibinfo {volume} {16}},\
  \bibinfo {pages} {4574} (\bibinfo {year} {2020})}\BibitemShut {NoStop}%
\bibitem [{\citenamefont {Shin}\ \emph {et~al.}(2003)\citenamefont {Shin},
  \citenamefont {Jo},\ and\ \citenamefont {Mikos}}]{shin2003biomimetic}%
  \BibitemOpen
  \bibfield  {author} {\bibinfo {author} {\bibfnamefont {H.}~\bibnamefont
  {Shin}}, \bibinfo {author} {\bibfnamefont {S.}~\bibnamefont {Jo}},\ and\
  \bibinfo {author} {\bibfnamefont {A.~G.}\ \bibnamefont {Mikos}},\ }\bibfield
  {title} {\bibinfo {title} {Biomimetic materials for tissue engineering},\
  }\href@noop {} {\bibfield  {journal} {\bibinfo  {journal} {Biomaterials}\
  }\textbf {\bibinfo {volume} {24}},\ \bibinfo {pages} {4353} (\bibinfo {year}
  {2003})}\BibitemShut {NoStop}%
\bibitem [{\citenamefont {Trivedi}\ \emph {et~al.}(2008)\citenamefont
  {Trivedi}, \citenamefont {Rahn}, \citenamefont {Kier},\ and\ \citenamefont
  {Walker}}]{trivedi2008soft}%
  \BibitemOpen
  \bibfield  {author} {\bibinfo {author} {\bibfnamefont {D.}~\bibnamefont
  {Trivedi}}, \bibinfo {author} {\bibfnamefont {C.~D.}\ \bibnamefont {Rahn}},
  \bibinfo {author} {\bibfnamefont {W.~M.}\ \bibnamefont {Kier}},\ and\
  \bibinfo {author} {\bibfnamefont {I.~D.}\ \bibnamefont {Walker}},\ }\bibfield
   {title} {\bibinfo {title} {Soft robotics: Biological inspiration, state of
  the art, and future research},\ }\href@noop {} {\bibfield  {journal}
  {\bibinfo  {journal} {Applied bionics and biomechanics}\ }\textbf {\bibinfo
  {volume} {5}},\ \bibinfo {pages} {99} (\bibinfo {year} {2008})}\BibitemShut
  {NoStop}%
\bibitem [{\citenamefont {Wallin}\ \emph {et~al.}(2018)\citenamefont {Wallin},
  \citenamefont {Pikul},\ and\ \citenamefont {Shepherd}}]{wallin20183d}%
  \BibitemOpen
  \bibfield  {author} {\bibinfo {author} {\bibfnamefont {T.}~\bibnamefont
  {Wallin}}, \bibinfo {author} {\bibfnamefont {J.}~\bibnamefont {Pikul}},\ and\
  \bibinfo {author} {\bibfnamefont {R.~F.}\ \bibnamefont {Shepherd}},\
  }\bibfield  {title} {\bibinfo {title} {3d printing of soft robotic systems},\
  }\href@noop {} {\bibfield  {journal} {\bibinfo  {journal} {Nature Reviews
  Materials}\ }\textbf {\bibinfo {volume} {3}},\ \bibinfo {pages} {84}
  (\bibinfo {year} {2018})}\BibitemShut {NoStop}%
\bibitem [{\citenamefont {Wang}\ and\ \citenamefont
  {Du}(2020)}]{wang2020covid}%
  \BibitemOpen
  \bibfield  {author} {\bibinfo {author} {\bibfnamefont {J.}~\bibnamefont
  {Wang}}\ and\ \bibinfo {author} {\bibfnamefont {G.}~\bibnamefont {Du}},\
  }\bibfield  {title} {\bibinfo {title} {Covid-19 may transmit through
  aerosol},\ }\href@noop {} {\bibfield  {journal} {\bibinfo  {journal} {Irish
  Journal of Medical Science}\ }\textbf {\bibinfo {volume} {189}},\ \bibinfo
  {pages} {1143} (\bibinfo {year} {2020})}\BibitemShut {NoStop}%
\bibitem [{\citenamefont {Morawska}(2005)}]{morawska2005droplet}%
  \BibitemOpen
  \bibfield  {author} {\bibinfo {author} {\bibfnamefont {L.}~\bibnamefont
  {Morawska}},\ }\bibfield  {title} {\bibinfo {title} {Droplet fate in indoor
  environments, or can we prevent the spread of infection?},\ }in\ \href@noop
  {} {\emph {\bibinfo {booktitle} {Indoor Air 2005: Proceedings of the 10th
  International Conference on Indoor Air Quality and Climate}}}\ (\bibinfo
  {organization} {Tsinghua University Press},\ \bibinfo {year} {2005})\
  \bibinfo {pages} {9--23}\BibitemShut {NoStop}%
\bibitem [{\citenamefont {Garvey}\ \emph {et~al.}(2023)\citenamefont {Garvey},
  \citenamefont {Williams}, \citenamefont {Gardiner}, \citenamefont {Ruston},
  \citenamefont {Wilkinson}, \citenamefont {Kiernan}, \citenamefont {Walker},\
  and\ \citenamefont {Holden}}]{garvey2023sink}%
  \BibitemOpen
  \bibfield  {author} {\bibinfo {author} {\bibfnamefont {M.}~\bibnamefont
  {Garvey}}, \bibinfo {author} {\bibfnamefont {N.}~\bibnamefont {Williams}},
  \bibinfo {author} {\bibfnamefont {A.}~\bibnamefont {Gardiner}}, \bibinfo
  {author} {\bibfnamefont {C.}~\bibnamefont {Ruston}}, \bibinfo {author}
  {\bibfnamefont {M.}~\bibnamefont {Wilkinson}}, \bibinfo {author}
  {\bibfnamefont {M.}~\bibnamefont {Kiernan}}, \bibinfo {author} {\bibfnamefont
  {J.}~\bibnamefont {Walker}},\ and\ \bibinfo {author} {\bibfnamefont
  {E.}~\bibnamefont {Holden}},\ }\bibfield  {title} {\bibinfo {title} {The sink
  splash zone},\ }\href@noop {} {\bibfield  {journal} {\bibinfo  {journal}
  {Journal of Hospital Infection}\ }\textbf {\bibinfo {volume} {135}},\
  \bibinfo {pages} {154} (\bibinfo {year} {2023})}\BibitemShut {NoStop}%
\bibitem [{\citenamefont {Hamidah}\ \emph {et~al.}(2023)\citenamefont
  {Hamidah}, \citenamefont {Nindito}, \citenamefont {Garib}, \citenamefont
  {Nuswantoro},\ and\ \citenamefont {Santoso}}]{hamidah2023spray}%
  \BibitemOpen
  \bibfield  {author} {\bibinfo {author} {\bibfnamefont {N.}~\bibnamefont
  {Hamidah}}, \bibinfo {author} {\bibfnamefont {D.}~\bibnamefont {Nindito}},
  \bibinfo {author} {\bibfnamefont {T.}~\bibnamefont {Garib}}, \bibinfo
  {author} {\bibfnamefont {W.}~\bibnamefont {Nuswantoro}},\ and\ \bibinfo
  {author} {\bibfnamefont {M.}~\bibnamefont {Santoso}},\ }\bibfield  {title}
  {\bibinfo {title} {Spray dipping system sinks hygienic and water saving},\
  }in\ \href@noop {} {\emph {\bibinfo {booktitle} {AIP Conference
  Proceedings}}},\ Vol.\ \bibinfo {volume} {\textbf{2629}}\ (\bibinfo
  {organization} {AIP Publishing},\ \bibinfo {year} {2023})\BibitemShut
  {NoStop}%
\bibitem [{\citenamefont {Thurairajah}\ \emph {et~al.}(2025)\citenamefont
  {Thurairajah}, \citenamefont {Song}, \citenamefont {Zhu}, \citenamefont
  {Shi}, \citenamefont {Barlow}, \citenamefont {Hurd},\ and\ \citenamefont
  {Pan}}]{thurairajah2025splash}%
  \BibitemOpen
  \bibfield  {author} {\bibinfo {author} {\bibfnamefont {K.}~\bibnamefont
  {Thurairajah}}, \bibinfo {author} {\bibfnamefont {X.}~\bibnamefont {Song}},
  \bibinfo {author} {\bibfnamefont {J.}~\bibnamefont {Zhu}}, \bibinfo {author}
  {\bibfnamefont {M.}~\bibnamefont {Shi}}, \bibinfo {author} {\bibfnamefont
  {E.~A.}\ \bibnamefont {Barlow}}, \bibinfo {author} {\bibfnamefont {R.~C.}\
  \bibnamefont {Hurd}},\ and\ \bibinfo {author} {\bibfnamefont
  {Z.}~\bibnamefont {Pan}},\ }\bibfield  {title} {\bibinfo {title} {Splash-free
  urinals for global sustainability and accessibility: Design through physics
  and differential equations},\ }\href@noop {} {\bibfield  {journal} {\bibinfo
  {journal} {PNAS nexus}\ }\textbf {\bibinfo {volume} {4}},\ \bibinfo {pages}
  {pgaf087} (\bibinfo {year} {2025})}\BibitemShut {NoStop}%
\bibitem [{\citenamefont {Newell}\ \emph {et~al.}(2010)\citenamefont {Newell},
  \citenamefont {Koopmans}, \citenamefont {Verhoef}, \citenamefont {Duizer},
  \citenamefont {Aidara-Kane}, \citenamefont {Sprong}, \citenamefont
  {Opsteegh}, \citenamefont {Langelaar}, \citenamefont {Threfall},
  \citenamefont {Scheutz} \emph {et~al.}}]{newell2010food}%
  \BibitemOpen
  \bibfield  {author} {\bibinfo {author} {\bibfnamefont {D.~G.}\ \bibnamefont
  {Newell}}, \bibinfo {author} {\bibfnamefont {M.}~\bibnamefont {Koopmans}},
  \bibinfo {author} {\bibfnamefont {L.}~\bibnamefont {Verhoef}}, \bibinfo
  {author} {\bibfnamefont {E.}~\bibnamefont {Duizer}}, \bibinfo {author}
  {\bibfnamefont {A.}~\bibnamefont {Aidara-Kane}}, \bibinfo {author}
  {\bibfnamefont {H.}~\bibnamefont {Sprong}}, \bibinfo {author} {\bibfnamefont
  {M.}~\bibnamefont {Opsteegh}}, \bibinfo {author} {\bibfnamefont
  {M.}~\bibnamefont {Langelaar}}, \bibinfo {author} {\bibfnamefont
  {J.}~\bibnamefont {Threfall}}, \bibinfo {author} {\bibfnamefont
  {F.}~\bibnamefont {Scheutz}}, \emph {et~al.},\ }\bibfield  {title} {\bibinfo
  {title} {Food-borne diseases—the challenges of 20 years ago still persist
  while new ones continue to emerge},\ }\href@noop {} {\bibfield  {journal}
  {\bibinfo  {journal} {International journal of food microbiology}\ }\textbf
  {\bibinfo {volume} {139}},\ \bibinfo {pages} {S3} (\bibinfo {year}
  {2010})}\BibitemShut {NoStop}%
\bibitem [{\citenamefont {Klontz}\ \emph {et~al.}(1995)\citenamefont {Klontz},
  \citenamefont {Timbo}, \citenamefont {Fein},\ and\ \citenamefont
  {Levy}}]{klontz1995prevalence}%
  \BibitemOpen
  \bibfield  {author} {\bibinfo {author} {\bibfnamefont {K.~C.}\ \bibnamefont
  {Klontz}}, \bibinfo {author} {\bibfnamefont {B.}~\bibnamefont {Timbo}},
  \bibinfo {author} {\bibfnamefont {S.}~\bibnamefont {Fein}},\ and\ \bibinfo
  {author} {\bibfnamefont {A.}~\bibnamefont {Levy}},\ }\bibfield  {title}
  {\bibinfo {title} {Prevalence of selected food consumption and preparation
  behaviors associated with increased risks of food-borne disease},\
  }\href@noop {} {\bibfield  {journal} {\bibinfo  {journal} {Journal of Food
  Protection}\ }\textbf {\bibinfo {volume} {58}},\ \bibinfo {pages} {927}
  (\bibinfo {year} {1995})}\BibitemShut {NoStop}%
\bibitem [{\citenamefont {Lai}\ \emph {et~al.}(2022)\citenamefont {Lai},
  \citenamefont {Tang}, \citenamefont {Wang}, \citenamefont {Ren},
  \citenamefont {Kong}, \citenamefont {Jiao},\ and\ \citenamefont
  {Huang}}]{lai2022handling}%
  \BibitemOpen
  \bibfield  {author} {\bibinfo {author} {\bibfnamefont {H.}~\bibnamefont
  {Lai}}, \bibinfo {author} {\bibfnamefont {Y.}~\bibnamefont {Tang}}, \bibinfo
  {author} {\bibfnamefont {Z.}~\bibnamefont {Wang}}, \bibinfo {author}
  {\bibfnamefont {F.}~\bibnamefont {Ren}}, \bibinfo {author} {\bibfnamefont
  {L.}~\bibnamefont {Kong}}, \bibinfo {author} {\bibfnamefont {X.}~\bibnamefont
  {Jiao}},\ and\ \bibinfo {author} {\bibfnamefont {J.}~\bibnamefont {Huang}},\
  }\bibfield  {title} {\bibinfo {title} {Handling practice as a critical point
  influencing the transmission route of campylobacter throughout a commercial
  restaurant kitchen in china},\ }\href@noop {} {\bibfield  {journal} {\bibinfo
   {journal} {Food Control}\ }\textbf {\bibinfo {volume} {139}},\ \bibinfo
  {pages} {109056} (\bibinfo {year} {2022})}\BibitemShut {NoStop}%
\bibitem [{\citenamefont {Lai}\ \emph {et~al.}(2023)\citenamefont {Lai},
  \citenamefont {Tang}, \citenamefont {Ren}, \citenamefont {Jiao},\ and\
  \citenamefont {Huang}}]{lai2023evaluation}%
  \BibitemOpen
  \bibfield  {author} {\bibinfo {author} {\bibfnamefont {H.}~\bibnamefont
  {Lai}}, \bibinfo {author} {\bibfnamefont {Y.}~\bibnamefont {Tang}}, \bibinfo
  {author} {\bibfnamefont {F.}~\bibnamefont {Ren}}, \bibinfo {author}
  {\bibfnamefont {X.-a.}\ \bibnamefont {Jiao}},\ and\ \bibinfo {author}
  {\bibfnamefont {J.}~\bibnamefont {Huang}},\ }\bibfield  {title} {\bibinfo
  {title} {Evaluation of hygiene practice for reducing campylobacter
  contamination on cutting boards and risks associated with chicken handling in
  kitchen environment},\ }\href@noop {} {\bibfield  {journal} {\bibinfo
  {journal} {Foods}\ }\textbf {\bibinfo {volume} {12}},\ \bibinfo {pages}
  {3245} (\bibinfo {year} {2023})}\BibitemShut {NoStop}%
\bibitem [{\citenamefont {Lefebvre}\ and\ \citenamefont
  {McDonell}(2017)}]{lefebvre2017atomization}%
  \BibitemOpen
  \bibfield  {author} {\bibinfo {author} {\bibfnamefont {A.~H.}\ \bibnamefont
  {Lefebvre}}\ and\ \bibinfo {author} {\bibfnamefont {V.~G.}\ \bibnamefont
  {McDonell}},\ }\href@noop {} {\emph {\bibinfo {title} {Atomization and
  sprays}}}\ (\bibinfo  {publisher} {CRC press, Boca Raton},\ \bibinfo {year}
  {2017})\BibitemShut {NoStop}%
\bibitem [{\citenamefont {Lin}\ and\ \citenamefont
  {Reitz}(1998)}]{lin1998drop}%
  \BibitemOpen
  \bibfield  {author} {\bibinfo {author} {\bibfnamefont {S.~P.}\ \bibnamefont
  {Lin}}\ and\ \bibinfo {author} {\bibfnamefont {R.~D.}\ \bibnamefont
  {Reitz}},\ }\bibfield  {title} {\bibinfo {title} {Drop and spray formation
  from a liquid jet},\ }\href@noop {} {\bibfield  {journal} {\bibinfo
  {journal} {Annual review of fluid mechanics}\ }\textbf {\bibinfo {volume}
  {30}},\ \bibinfo {pages} {85} (\bibinfo {year} {1998})}\BibitemShut {NoStop}%
\bibitem [{\citenamefont {Dumouchel}(2008)}]{dumouchel2008experimental}%
  \BibitemOpen
  \bibfield  {author} {\bibinfo {author} {\bibfnamefont {C.}~\bibnamefont
  {Dumouchel}},\ }\bibfield  {title} {\bibinfo {title} {On the experimental
  investigation on primary atomization of liquid streams},\ }\href@noop {}
  {\bibfield  {journal} {\bibinfo  {journal} {Experiments in fluids}\ }\textbf
  {\bibinfo {volume} {45}},\ \bibinfo {pages} {371} (\bibinfo {year}
  {2008})}\BibitemShut {NoStop}%
\bibitem [{\citenamefont {Kim}\ \emph {et~al.}(2019)\citenamefont {Kim},
  \citenamefont {Park}, \citenamefont {Gruszewski}, \citenamefont {Schmale},\
  and\ \citenamefont {Jung}}]{kim2019vortex}%
  \BibitemOpen
  \bibfield  {author} {\bibinfo {author} {\bibfnamefont {S.}~\bibnamefont
  {Kim}}, \bibinfo {author} {\bibfnamefont {H.}~\bibnamefont {Park}}, \bibinfo
  {author} {\bibfnamefont {H.~A.}\ \bibnamefont {Gruszewski}}, \bibinfo
  {author} {\bibfnamefont {D.~G.}\ \bibnamefont {Schmale}},\ and\ \bibinfo
  {author} {\bibfnamefont {S.}~\bibnamefont {Jung}},\ }\bibfield  {title}
  {\bibinfo {title} {Vortex-induced dispersal of a plant pathogen by raindrop
  impact},\ }\href@noop {} {\bibfield  {journal} {\bibinfo  {journal}
  {Proceedings of the National Academy of Sciences}\ }\textbf {\bibinfo
  {volume} {116}},\ \bibinfo {pages} {4917} (\bibinfo {year}
  {2019})}\BibitemShut {NoStop}%
\bibitem [{\citenamefont {Clift}\ and\ \citenamefont
  {Gauvin}(1971)}]{clift1971motion}%
  \BibitemOpen
  \bibfield  {author} {\bibinfo {author} {\bibfnamefont {R.}~\bibnamefont
  {Clift}}\ and\ \bibinfo {author} {\bibfnamefont {W.}~\bibnamefont {Gauvin}},\
  }\bibfield  {title} {\bibinfo {title} {Motion of entrained particles in gas
  streams},\ }\href@noop {} {\bibfield  {journal} {\bibinfo  {journal} {The
  Canadian Journal of Chemical Engineering}\ }\textbf {\bibinfo {volume}
  {49}},\ \bibinfo {pages} {439} (\bibinfo {year} {1971})}\BibitemShut
  {NoStop}%
\bibitem [{\citenamefont {Hasegawa}\ \emph {et~al.}(1997)\citenamefont
  {Hasegawa}, \citenamefont {Suganuma},\ and\ \citenamefont
  {Watanabe}}]{hasegawa1997anomaly}%
  \BibitemOpen
  \bibfield  {author} {\bibinfo {author} {\bibfnamefont {T.}~\bibnamefont
  {Hasegawa}}, \bibinfo {author} {\bibfnamefont {M.}~\bibnamefont {Suganuma}},\
  and\ \bibinfo {author} {\bibfnamefont {H.}~\bibnamefont {Watanabe}},\
  }\bibfield  {title} {\bibinfo {title} {Anomaly of excess pressure drops of
  the flow through very small orifices},\ }\href@noop {} {\bibfield  {journal}
  {\bibinfo  {journal} {Physics of Fluids}\ }\textbf {\bibinfo {volume} {9}},\
  \bibinfo {pages} {1} (\bibinfo {year} {1997})}\BibitemShut {NoStop}%
\bibitem [{\citenamefont {Villermaux}(2007)}]{villermaux2007fragmentation}%
  \BibitemOpen
  \bibfield  {author} {\bibinfo {author} {\bibfnamefont {E.}~\bibnamefont
  {Villermaux}},\ }\bibfield  {title} {\bibinfo {title} {Fragmentation},\
  }\href@noop {} {\bibfield  {journal} {\bibinfo  {journal} {Annu. Rev. Fluid
  Mech.}\ }\textbf {\bibinfo {volume} {39}},\ \bibinfo {pages} {419} (\bibinfo
  {year} {2007})}\BibitemShut {NoStop}%
\bibitem [{\citenamefont {Wang}\ and\ \citenamefont
  {Bourouiba}(2018)}]{wang2018unsteady}%
  \BibitemOpen
  \bibfield  {author} {\bibinfo {author} {\bibfnamefont {Y.}~\bibnamefont
  {Wang}}\ and\ \bibinfo {author} {\bibfnamefont {L.}~\bibnamefont
  {Bourouiba}},\ }\bibfield  {title} {\bibinfo {title} {Unsteady sheet
  fragmentation: droplet sizes and speeds},\ }\href@noop {} {\bibfield
  {journal} {\bibinfo  {journal} {Journal of Fluid Mechanics}\ }\textbf
  {\bibinfo {volume} {848}},\ \bibinfo {pages} {946} (\bibinfo {year}
  {2018})}\BibitemShut {NoStop}%
\bibitem [{\citenamefont {Yamagata}\ \emph {et~al.}(2017)\citenamefont
  {Yamagata}, \citenamefont {Lu}, \citenamefont {Sekiguchi},\ and\
  \citenamefont {Sato}}]{yamagata2017experimental}%
  \BibitemOpen
  \bibfield  {author} {\bibinfo {author} {\bibfnamefont {Y.}~\bibnamefont
  {Yamagata}}, \bibinfo {author} {\bibfnamefont {X.}~\bibnamefont {Lu}},
  \bibinfo {author} {\bibfnamefont {Y.}~\bibnamefont {Sekiguchi}},\ and\
  \bibinfo {author} {\bibfnamefont {C.}~\bibnamefont {Sato}},\ }\bibfield
  {title} {\bibinfo {title} {Experimental investigation of mode i fracture
  energy of adhesively bonded joints under impact loading conditions},\
  }\href@noop {} {\bibfield  {journal} {\bibinfo  {journal} {Applied adhesion
  science}\ }\textbf {\bibinfo {volume} {5}},\ \bibinfo {pages} {1} (\bibinfo
  {year} {2017})}\BibitemShut {NoStop}%
\bibitem [{\citenamefont {Mahiuddin}\ \emph {et~al.}(2020)\citenamefont
  {Mahiuddin}, \citenamefont {Godhani}, \citenamefont {Feng}, \citenamefont
  {Liu}, \citenamefont {Langrish},\ and\ \citenamefont
  {Karim}}]{mahiuddin2020application}%
  \BibitemOpen
  \bibfield  {author} {\bibinfo {author} {\bibfnamefont {M.}~\bibnamefont
  {Mahiuddin}}, \bibinfo {author} {\bibfnamefont {D.}~\bibnamefont {Godhani}},
  \bibinfo {author} {\bibfnamefont {L.}~\bibnamefont {Feng}}, \bibinfo {author}
  {\bibfnamefont {F.}~\bibnamefont {Liu}}, \bibinfo {author} {\bibfnamefont
  {T.}~\bibnamefont {Langrish}},\ and\ \bibinfo {author} {\bibfnamefont
  {M.}~\bibnamefont {Karim}},\ }\bibfield  {title} {\bibinfo {title}
  {Application of caputo fractional rheological model to determine the
  viscoelastic and mechanical properties of fruit and vegetables},\ }\href@noop
  {} {\bibfield  {journal} {\bibinfo  {journal} {Postharvest Biology and
  Technology}\ }\textbf {\bibinfo {volume} {163}},\ \bibinfo {pages} {111147}
  (\bibinfo {year} {2020})}\BibitemShut {NoStop}%
\bibitem [{\citenamefont {Johnson}\ and\ \citenamefont
  {Johnson}(1987)}]{johnson1987contact}%
  \BibitemOpen
  \bibfield  {author} {\bibinfo {author} {\bibfnamefont {K.~L.}\ \bibnamefont
  {Johnson}}\ and\ \bibinfo {author} {\bibfnamefont {K.~L.}\ \bibnamefont
  {Johnson}},\ }\href@noop {} {\emph {\bibinfo {title} {Contact mechanics}}}\
  (\bibinfo  {publisher} {Cambridge university press, Cambridge},\ \bibinfo
  {year} {1987})\BibitemShut {NoStop}%
\bibitem [{\citenamefont {Sneddon}(1965)}]{sneddon1965relation}%
  \BibitemOpen
  \bibfield  {author} {\bibinfo {author} {\bibfnamefont {I.~N.}\ \bibnamefont
  {Sneddon}},\ }\bibfield  {title} {\bibinfo {title} {The relation between load
  and penetration in the axisymmetric boussinesq problem for a punch of
  arbitrary profile},\ }\href@noop {} {\bibfield  {journal} {\bibinfo
  {journal} {International journal of engineering science}\ }\textbf {\bibinfo
  {volume} {3}},\ \bibinfo {pages} {47} (\bibinfo {year} {1965})}\BibitemShut
  {NoStop}%
\bibitem [{\citenamefont {Jafari~Malekabadi}\ \emph {et~al.}(2016)\citenamefont
  {Jafari~Malekabadi}, \citenamefont {Khojastehpour}, \citenamefont {Emadi},\
  and\ \citenamefont {Golzarian}}]{jafari2016determination}%
  \BibitemOpen
  \bibfield  {author} {\bibinfo {author} {\bibfnamefont {A.}~\bibnamefont
  {Jafari~Malekabadi}}, \bibinfo {author} {\bibfnamefont {M.}~\bibnamefont
  {Khojastehpour}}, \bibinfo {author} {\bibfnamefont {B.}~\bibnamefont
  {Emadi}},\ and\ \bibinfo {author} {\bibfnamefont {M.}~\bibnamefont
  {Golzarian}},\ }\bibfield  {title} {\bibinfo {title} {Determination of
  elasticity modulus and poisson ratio of two onion varieties under different
  loading conditions},\ }\href@noop {} {\bibfield  {journal} {\bibinfo
  {journal} {Journal of Agricultural Machinery}\ }\textbf {\bibinfo {volume}
  {6}},\ \bibinfo {pages} {114} (\bibinfo {year} {2016})}\BibitemShut {NoStop}%
\bibitem [{\citenamefont {Zhu}\ and\ \citenamefont
  {Melrose}(2003)}]{zhu2003mechanics}%
  \BibitemOpen
  \bibfield  {author} {\bibinfo {author} {\bibfnamefont {H.}~\bibnamefont
  {Zhu}}\ and\ \bibinfo {author} {\bibfnamefont {J.}~\bibnamefont {Melrose}},\
  }\bibfield  {title} {\bibinfo {title} {A mechanics model for the compression
  of plant and vegetative tissues},\ }\href@noop {} {\bibfield  {journal}
  {\bibinfo  {journal} {Journal of theoretical biology}\ }\textbf {\bibinfo
  {volume} {221}},\ \bibinfo {pages} {89} (\bibinfo {year} {2003})}\BibitemShut
  {NoStop}%
\bibitem [{\citenamefont {Synge}\ and\ \citenamefont
  {Griffith}(2011)}]{synge2011principles}%
  \BibitemOpen
  \bibfield  {author} {\bibinfo {author} {\bibfnamefont {J.~L.}\ \bibnamefont
  {Synge}}\ and\ \bibinfo {author} {\bibfnamefont {B.~A.}\ \bibnamefont
  {Griffith}},\ }\href@noop {} {\emph {\bibinfo {title} {Principles of
  Mechanics, 2nd Edition}}}\ (\bibinfo  {publisher} {Mcgraw-Hill Book Company,
  New York},\ \bibinfo {year} {2011})\
  \bibinfo {pages}
  {99--105}\BibitemShut {NoStop}%
\bibitem [{\citenamefont {Vincent}(1990)}]{vincent1990fracture}%
  \BibitemOpen
  \bibfield  {author} {\bibinfo {author} {\bibfnamefont {J.~F.}\ \bibnamefont
  {Vincent}},\ }\bibfield  {title} {\bibinfo {title} {Fracture properties of
  plants},\ }in\ \href@noop {} {\emph {\bibinfo {booktitle} {Advances in
  botanical research}}},\ Vol.\ \bibinfo {volume} {\textbf{17}}\ (\bibinfo
  {publisher} {Elsevier},\ \bibinfo {year} {1990})\
  \bibinfo {pages}
  {235--287}\BibitemShut {NoStop}%
\bibitem [{\citenamefont {Bidhendi}\ \emph {et~al.}(2023)\citenamefont
  {Bidhendi}, \citenamefont {Lampron}, \citenamefont {Gosselin},\ and\
  \citenamefont {Geitmann}}]{bidhendi2023cell}%
  \BibitemOpen
  \bibfield  {author} {\bibinfo {author} {\bibfnamefont {A.~J.}\ \bibnamefont
  {Bidhendi}}, \bibinfo {author} {\bibfnamefont {O.}~\bibnamefont {Lampron}},
  \bibinfo {author} {\bibfnamefont {F.~P.}\ \bibnamefont {Gosselin}},\ and\
  \bibinfo {author} {\bibfnamefont {A.}~\bibnamefont {Geitmann}},\ }\bibfield
  {title} {\bibinfo {title} {Cell geometry regulates tissue fracture},\
  }\href@noop {} {\bibfield  {journal} {\bibinfo  {journal} {Nature
  Communications}\ }\textbf {\bibinfo {volume} {\textbf{14}}},\ \bibinfo
  {pages} {8275} (\bibinfo {year} {2023})}\BibitemShut {NoStop}%
\bibitem [{Epi()}]{Epicurious}%
  \BibitemOpen
  \href@noop {} {\bibinfo {title} {How to actually prevent tears when chopping
  onions}},\ \bibinfo {howpublished}
  {https://www.epicurious.com/expert-advice/how-to-cut-onions-without-crying}\BibitemShut
  {NoStop}%
\bibitem [{\citenamefont {Treitel}(1946)}]{treitel1946elasticity}%
  \BibitemOpen
  \bibfield  {author} {\bibinfo {author} {\bibfnamefont {O.}~\bibnamefont
  {Treitel}},\ }\bibfield  {title} {\bibinfo {title} {Elasticity, plasticity
  and fine structure of plant cell walls},\ }\href@noop {} {\bibfield
  {journal} {\bibinfo  {journal} {Journal of Colloid Science}\ }\textbf
  {\bibinfo {volume} {1}},\ \bibinfo {pages} {327} (\bibinfo {year}
  {1946})}\BibitemShut {NoStop}%
\bibitem [{\citenamefont {Berger}\ \emph {et~al.}(2010)\citenamefont {Berger},
  \citenamefont {Sodha}, \citenamefont {Shaw}, \citenamefont {Griffin},
  \citenamefont {Pink}, \citenamefont {Hand},\ and\ \citenamefont
  {Frankel}}]{berger2010fresh}%
  \BibitemOpen
  \bibfield  {author} {\bibinfo {author} {\bibfnamefont {C.~N.}\ \bibnamefont
  {Berger}}, \bibinfo {author} {\bibfnamefont {S.~V.}\ \bibnamefont {Sodha}},
  \bibinfo {author} {\bibfnamefont {R.~K.}\ \bibnamefont {Shaw}}, \bibinfo
  {author} {\bibfnamefont {P.~M.}\ \bibnamefont {Griffin}}, \bibinfo {author}
  {\bibfnamefont {D.}~\bibnamefont {Pink}}, \bibinfo {author} {\bibfnamefont
  {P.}~\bibnamefont {Hand}},\ and\ \bibinfo {author} {\bibfnamefont
  {G.}~\bibnamefont {Frankel}},\ }\bibfield  {title} {\bibinfo {title} {Fresh
  fruit and vegetables as vehicles for the transmission of human pathogens},\
  }\href@noop {} {\bibfield  {journal} {\bibinfo  {journal} {Environmental
  microbiology}\ }\textbf {\bibinfo {volume} {12}},\ \bibinfo {pages} {2385}
  (\bibinfo {year} {2010})}\BibitemShut {NoStop}%
\end{thebibliography}

%%%%%%%%%%%%%%%%%%%%%%%%%%%%%%%%%%%%%%

%%%%%%%%%%%%%%%%%%%%%%%%%%%%%%%%%%%%%%%

\end{document}